\documentclass[%
 preprint,
 amsmath,amssymb,
 aps,
]{revtex4-1}

\usepackage{graphicx}
\usepackage{dcolumn}
\usepackage{bm}
\usepackage[draft]{hyperref}
\usepackage{amsmath}
\usepackage{textcomp}
\usepackage{float}
\usepackage{subfig}
\usepackage{wrapfig}
\usepackage{fancyhdr}
\floatstyle{boxed}
\begin{document}

\title{Self assembly of microparticles in stable ring structures in an optical trap}


\author{Arijit Haldar}
 \affiliation{Department of Physical Sciences, IISER-Kolkata, Mohanpur 741252, India}
\author{Sambit Bikas Pal}%
\affiliation{Department of Physical Sciences, IISER-Kolkata, Mohanpur 741252, India}%
\author{Basudev Roy}
\affiliation{Department of Physical Sciences, IISER-Kolkata, Mohanpur 741252, India}
\author{S. Dutta Gupta}
\affiliation{School of Physics, University of Hyderabad, Hyderabad 500046, India}
\author{Ayan Banerjee}
 \email{ayan@iiserkol.ac.in}
\affiliation{Department of Physical Sciences, IISER-Kolkata, Mohanpur 741252, India}\thanks{}
\date{\today}
\begin{abstract}
Micro-particle self assembly under the influence of optical forces produced by higher order optical beams or by projection of a hologram into the trapping volume is well known. In this paper, we report the spontaneous formation of a ring of identical microspheres (each with diameter 1.1 $\mu$m) in conventional single beam optical tweezers with a usual TEM$_{00}$ Gaussian beam coupled into a sample chamber having standing wave geometry with a cover slip and glass slide. The effects of different experimental parameters on the ring formation are studied extensively. The experimental observations are backed by theoretical simulations based on a plane wave decomposition of the forward and backward propagating Gaussian beams. The ring patterns are shown to be caused due to geomterical aberrations produced by focusing the Gaussian beam using a high numerical aperture microscope objective into stratified media. It is found that the thickness of the stratified media and the standing wave geometry itself play a critical role in formation of stable ring structures. These structures could be used in the study of optical binding, as well as biological interactions between cells in an optical trap.
\begin{description}
\item[PACS numbers: 42.50.Wk, 87.80.Cc, 42.25.-p]
\end{description}

\end{abstract}

\maketitle
\section{\label{intro}INTRODUCTION}
\noindent Optical tweezers offer the possibilities of applying controlled forces of extremely small magnitude (pN-fN) to trap and manipulate mesoscopic particles  - as a result of which they have widespread applications in physics \cite{Ghi94, volpe06, clapp99} and biology \cite{svo94, mehta99, smith01, wen07}.  For relatively large particles (size between few to several tens of microns), it is easy to trap single particles at the trap center, while for nanoparticles, clusters assemble with time in the central region \cite{Hoso05}. Micro-particle self-assembly is interesting since it provides an insight into the dynamics of particles in a well-understood force environment, interactions between particles (such as cells \cite{Old83}), and processes where particles themselves modify the force environment as is manifested in optical binding \cite{Dhol10}. The most widely used way of creating microparticle assembly in optical traps is by modifying the trapping beam to include higher order axial modes (TEM$_{10}$, TEM$_{11}$, etc), vortex beams (Laguerre-Gaussian modes), as well as interference patterns by using a spatial light modulator \cite{Pad11}. Micro-particle assembly has also been observed due to surface tension effects between adjacent particles. It has been shown that multiple polystyrene micro-spheres can coalesce together inside a single trap, to form highly symmetric closed pack structures \cite{Lee06}. Another effect that has shown to form specifically ring-like structures at the laser focus is the phenomenon of thermo-capillarity \cite{Ahl07}, where the high power of the trapping laser causes convection currents at the focal spot, resulting in the micro-spheres being displaced away from the center. When an equilibrium is reached between the trapping force and the forces exerted from the convection currents, a ring like structure forms. Thermo-capillary effects are highly sensitive to the change in the power of the trapping beam, with the ring diameter increasing proportionally with the power until the chain of particles breaks at certain powers.

In this paper, we report the stable trapping of  polystyrene beads of diameter 1.1 $\mu$m in ring structures with the use a single pure Gaussian beam (TEM$_{00}$) for trapping the particles. The diameter of the rings can be varied between $\sim 3-5~ \mu$m by changing the z-focus of the trapping microscope, with the number of trapped particles in a closed ring varying between 9-15. Our sample chamber consisted of a cover slip and a slide, with the sample (polystyrene beads dispersed in water) sandwiched between. This particular configuration was critical to the formation of rings, which were not seen in the absence of a top slide. Our beam was checked to be a pure Gaussian TEM$_{00}$ by a standard beam profile measurement using a sharp edge mounted on a translation stage - in addition, to remove any uncertainty about higher order axial modes being contained in the beam, we coupled our trapping laser into a single mode optical fiber. The ring formation remained unaffected by this step in the experimental procedure, which thus implied that the effect was not due to such higher order modes in the trapping beam itself. Also, the closed pack structures reported in Ref. [\onlinecite{Lee06}] are not dependent on the presence of a top slide, and typically occur when one of the beads is in {\it contact} with a surface - both circumstances being very different from our observations. Such dependence is not observed in our experiments. In addition, the fact that we do not observe any dependence of ring radius on the power of the laser beam indicates that thermo-capillary effects \cite{Ahl07} are not involved in the formation of particle rings. 

It therefore appears that the phenomenon we observe has not been reported in literature. We thus perform theoretical simulations to understand the structure of the electric field inside the sample chamber that could support the formation of such stable ring structures. The structure of the paper is as follows. In Section~\ref{expmt}, we discuss the experimental conditions for obtaining the ring structure in detail. In Section~\ref{analys}, we lay out the theoretical model we have used to find the field in both radial and axial directions inside the sample chamber. Section~\ref{resdis} describes the simulations performed and a thorough analysis of the simulation results to explain our experimental findings. We end the paper with a few possible applications of such ring formation and future work that is planned in Section~\ref{concl}.

\section{\label{expmt}EXPERIMENT AND OBSERVATIONS}
The setup consisted of an inverted microscope with a high power IR laser (Lasever  LSR1064ML, 1064 nm, max power 800 mW) fed into the microscope (Carl Zeiss Axiovert.A1) back port, so that the laser beam was focused tightly on the sample using a high numerical aperture objective (Zeiss 100X, oil immersion, plan apochromat, 1.41 NA, infinity corrected) lens arrangement. The typical specified working distance of this objective is 170 $\mu$m - however, we observed that we could focus into our sample chamber at distances around 300 $\mu$m from the exit pupil of the objective. This was verified by using the 250 $\mu$m cover slip and putting two fiducial marks - one on the top surface of the cover slip, and the other on the bottom surface of the slide. We observed that when we increased the sample thickness to more than 40 $\mu$m, the fiducial mark on the slide could no longer be imaged. The sample consists of a dilute solution of monodispersed polystyrene micro-spheres (mean diameter 1.1 $\mu$m, Sigma Aldrich LB11) in distilled water (dilution 1:10000) placed in a sample chamber consisting of a glass slide and cover slip. The cover slips we used were of two types: 1) made of English glass (RI = 1.512 at 1064 nm) having thickness 160 $\mu$m, and 2) made of a polymer (Sigma Aldrich Hybrid Cover-slips,
Part no. Z365912-100EA) having refractive index of 1.575 (at 1064 nm) and thickness of 250 $\mu$m. The refractive index of the Type 2 cover slips was measured using a commercial refractometer (Prism Coupler - Metricon Model-2010). Therefore, only in the case of the 250 $\mu$m cover slip was there a significant refractive index mismatch between the cover slip and immersion oil.
We used two methods of preparing the sample chamber - a) Configuration 1, where we put the sample on the cover slip and directly attached the cover slip on top so that there was a thin film of micro-sphere solution between slide and cover slip, and b) Configuration 2, where we used a spacer of thickness around 100 $\mu$m consisting of double-sided sticky tape between the cover slip and slide. Note that this configuration could not be used for 250 $\mu$m cover slips due to focusing problems of the microscope objective. 
\begin{figure}[!h!t]
    \centering
  \subfloat[Configuration 1: Sample chamber without spacer]
{\includegraphics[scale=0.4]{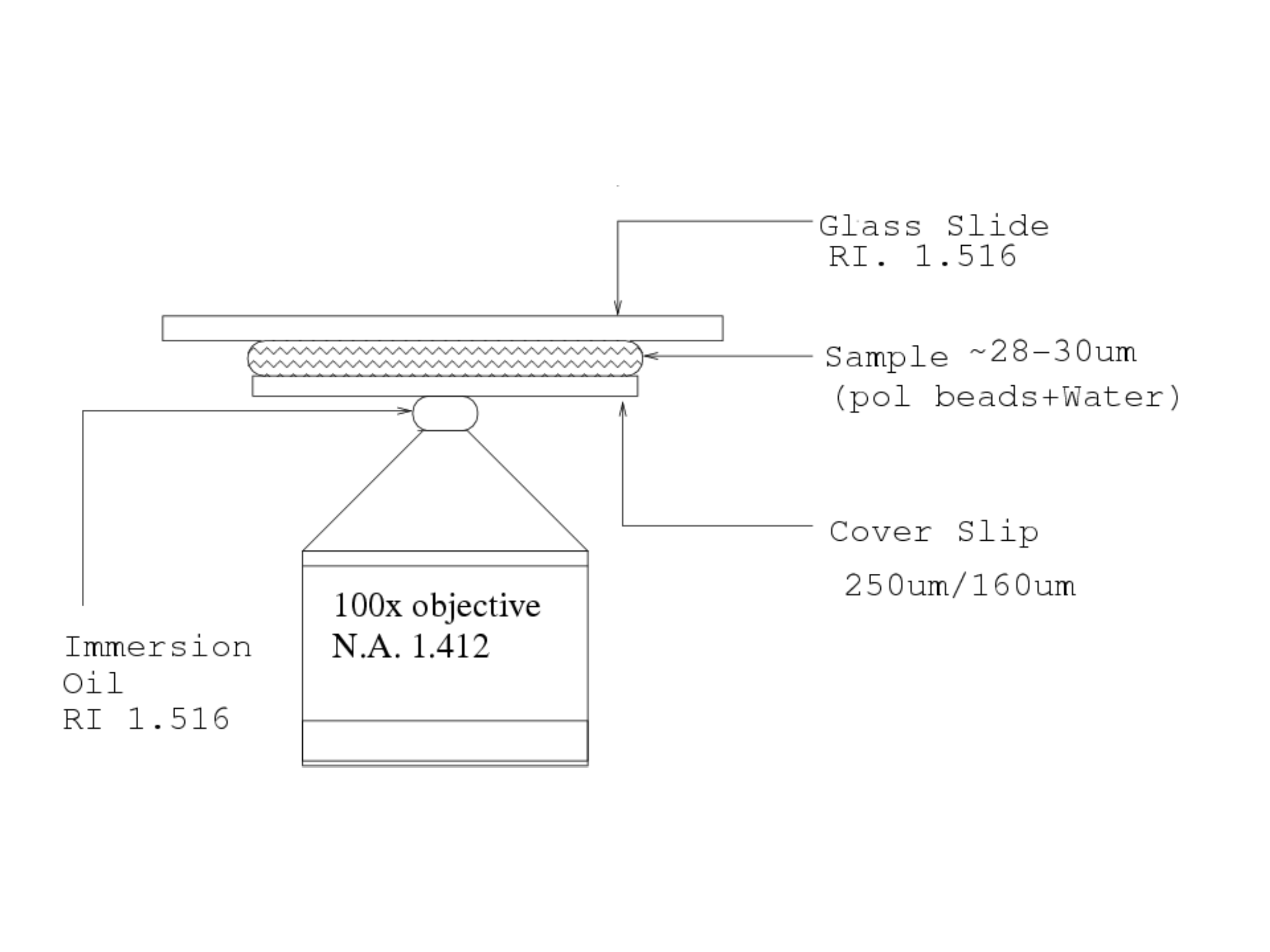}}\label{setup2}
   \subfloat[Configuration 2: Sample chamber with spacer] {\includegraphics[scale=0.4]{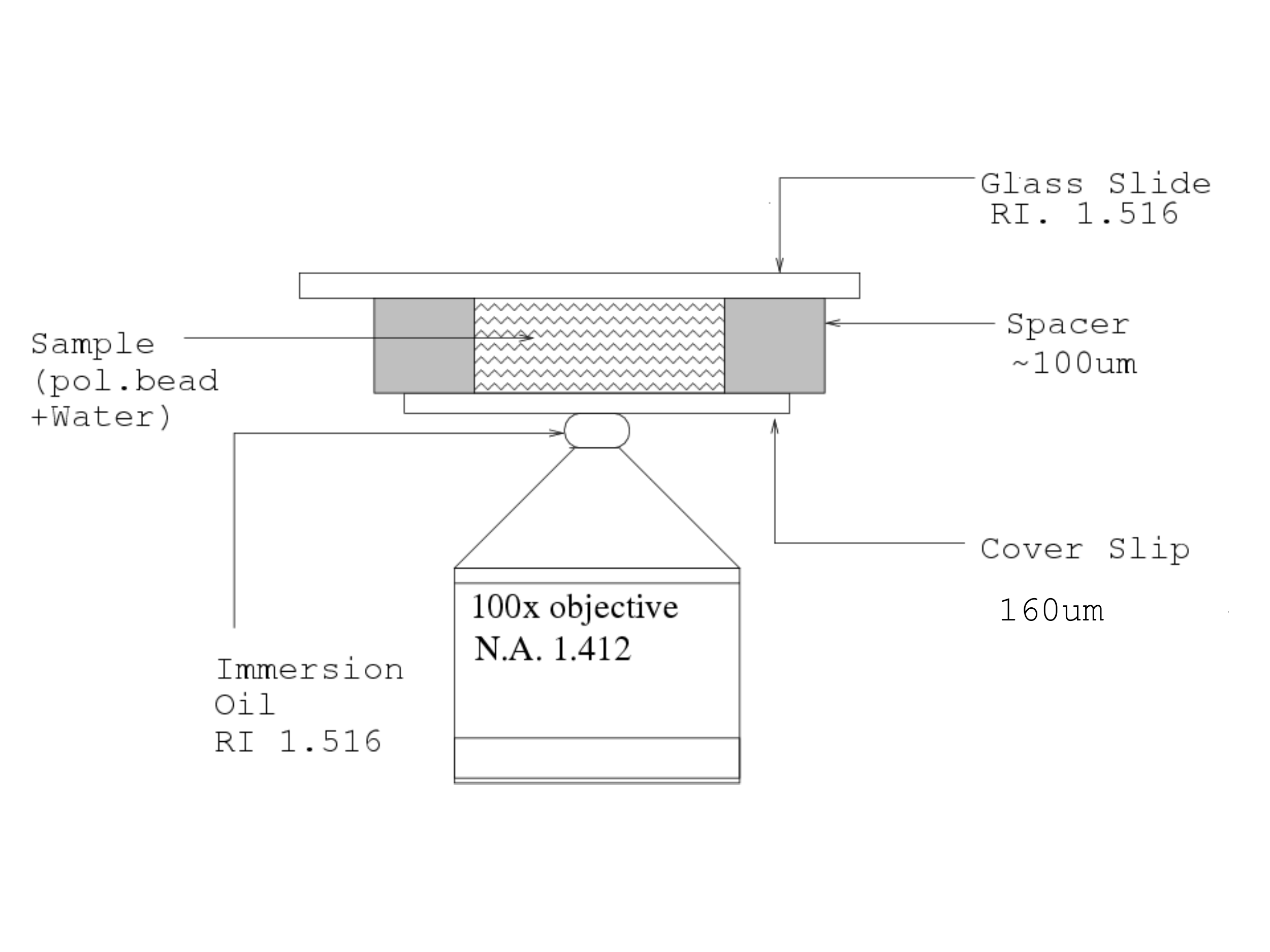}}\label{setup3}
     \caption{Two types of sample chamber configurations used in the experiment}
 \label{samplechamb}
   \end{figure}
The arrangements are shown in Fig.~\ref{samplechamb}.

\begin{figure}[!h!t]
\centering
  \subfloat[]{\includegraphics[height=100pt,width=100pt]{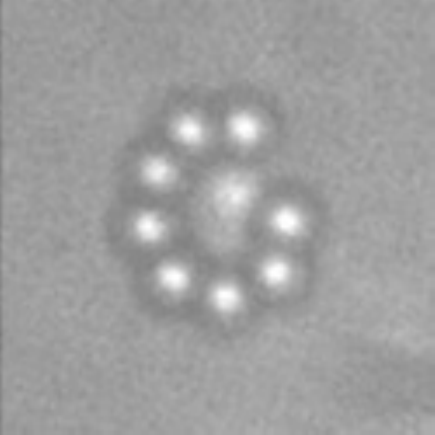}\label{ring3}}
 \subfloat[]{\includegraphics[height=100pt,width=100pt]{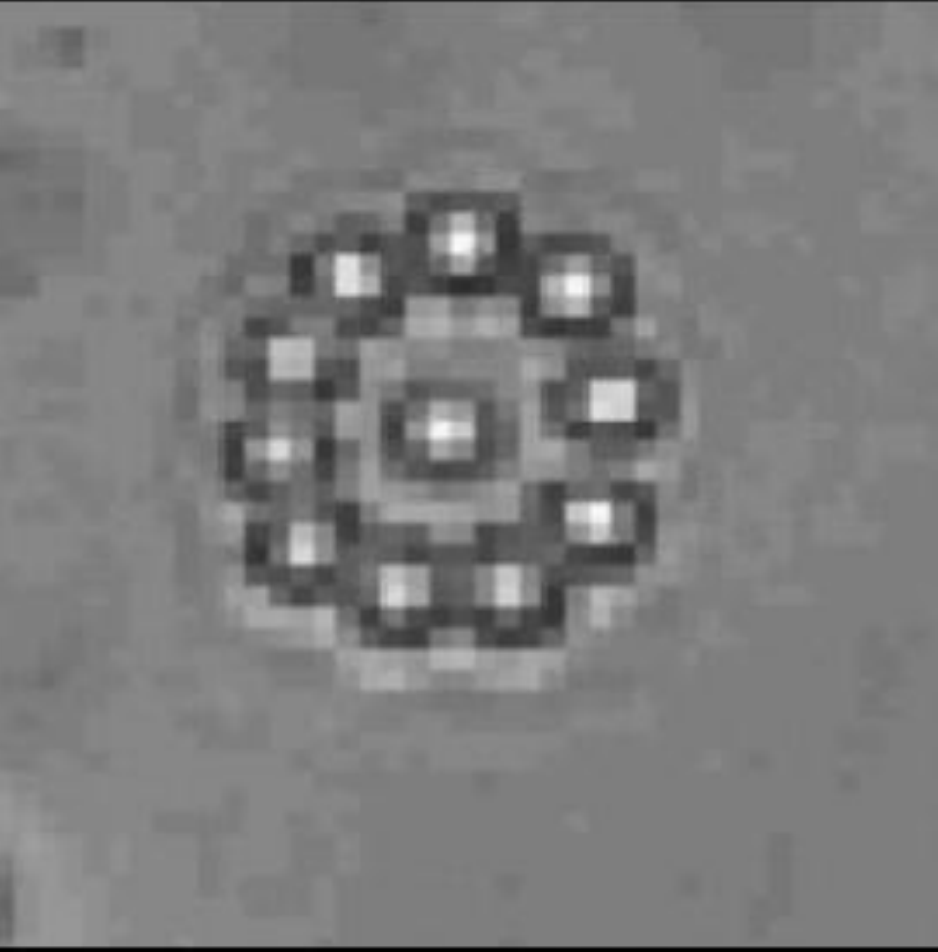}\label{ring4}}\\
  \subfloat[]{\includegraphics[height=100pt,width=100pt]{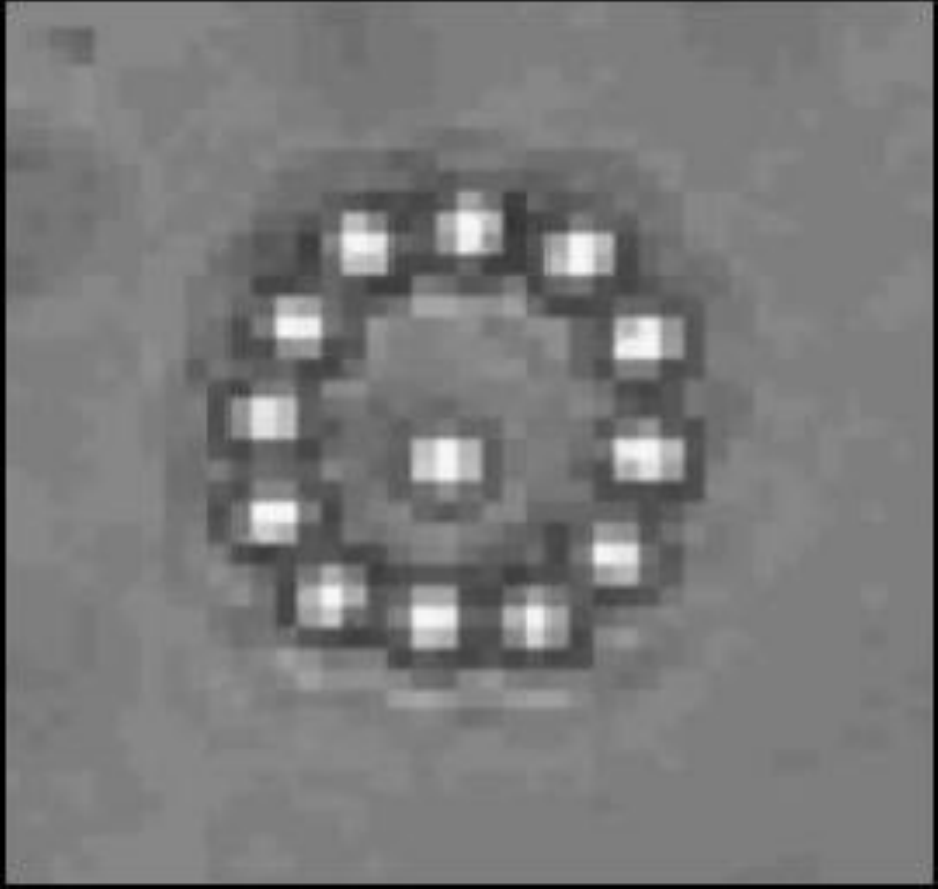}\label{ring5}}
 \subfloat[]{\includegraphics[height=100pt,width=100pt]{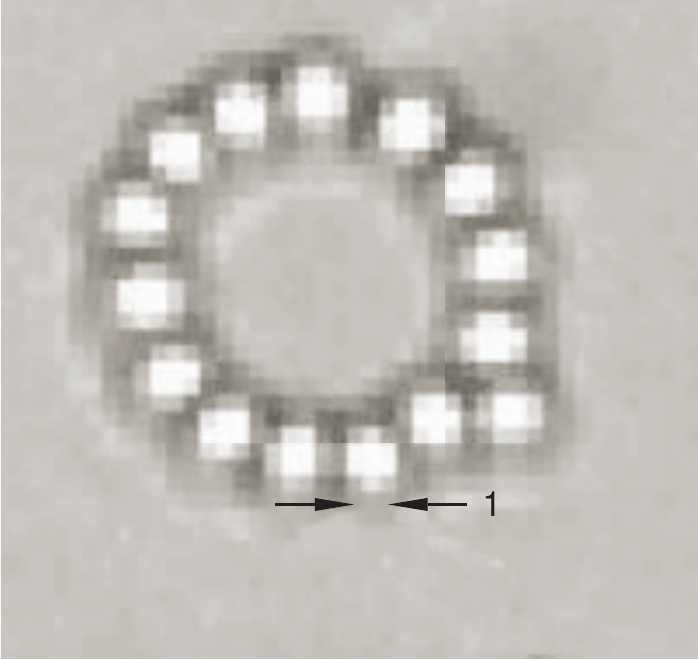}\label{ring6}}
  \caption{\label{radialpattern1}Radial Pattern formation with 250 $\mu$m cover slips with (a) 8, (b)  10, (c) 12, and (d) 14 particles in the ring. (a), (b), and (c) could possibly accommodate one more particle. Dimensions are in microns.}
\end{figure}

When we performed the experiment using configuration 1 with 250 $\mu$m thick cover slips, for a sample chamber thickness of around 30 $\mu$m (distance between slide and cover slip), we observed a spontaneous formation of ring-like structures of microparticles as shown in Fig.~\ref{radialpattern1}. The particles were allowed to accumulate gradually, and soon formed closed ring structures after slight manipulation of the microscope focus. If the trap was switched off, the particles diffused away, only to reassemble almost instantaneously in the same ring structure after the trap was switched on again. There was occasionally a particle trapped in the center, but the most likely region of trapping was in the ring some distance away from the center (also the focus of the microscope). The radius of the ring could be varied as well by changing the focus of the microscope. We could change the diameter of the rings between a minimum of around 3 $\mu$m to 5 $\mu$m. The diameter was obtained by noting the circumference of the ring, which to close approximation was the product of the number of microspheres (that varied between 9 to 15) in a complete ring and the individual microsphere diameter (1.1 $\mu$m). The ring diameter could then be obtained by dividing the circumference by $\pi$. The diameter was also verified by analyzing the image of a given ring in standard softwares such as Canvas or Adobe Illustrator.

\begin{figure}[ht]
 \centering{\includegraphics[scale=0.45]{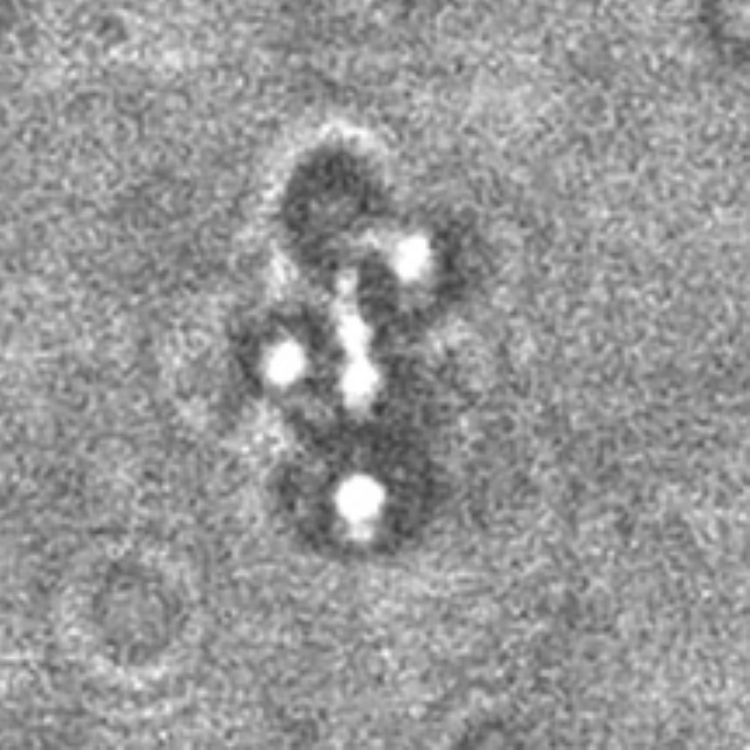}}
    \caption[clumping]{A cluster of 1.1 $\mu$m diameter particles trapped in the focus of the trapping laser for 160 $\mu$m thickness cover slips.}
\label{clumping}
\end{figure}

\begin{figure}[!h!t]
\centering
 \subfloat[]{\includegraphics[scale=0.5]{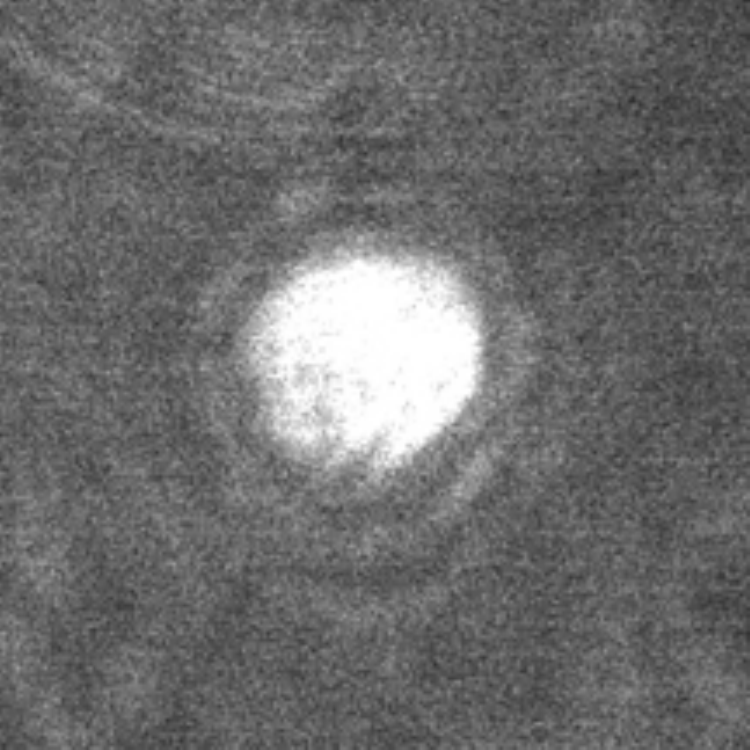}\label{160ring}}
\subfloat[]{\includegraphics[scale=0.5]{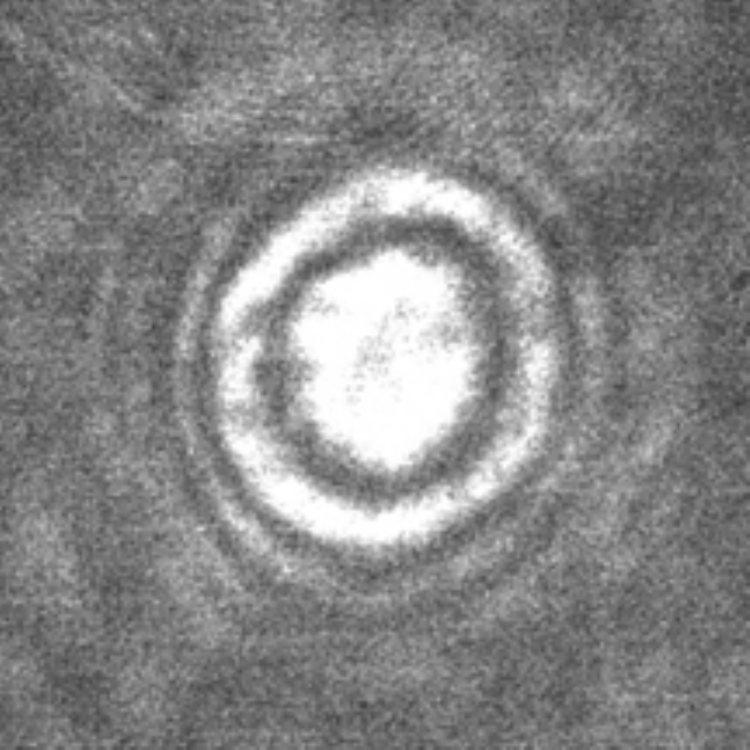}\label{250ring}} \caption{\label{nopartring}CCD camera images of the trapping field taken in the back focal plane of the trapping microscope for (a) 160 $\mu$m cover slips, and (b) 250 $\mu$m cover slips. The beam intensity in Case (a) is concentrated at the center with weak Airy rings visible as is expected for a focused Gaussian beam, but in Case (b), a ring structure is clearly seen with intensity similar to the central region. Note that the field pattern displayed in these images will be a superposition of the fields at all planes in the sample, containing both scattered fields from the micro-sphere, and reflected fields from the different chamber surfaces.}
\end{figure}

\begin{figure}[!h!t]
 \centering{\includegraphics[scale=0.5]{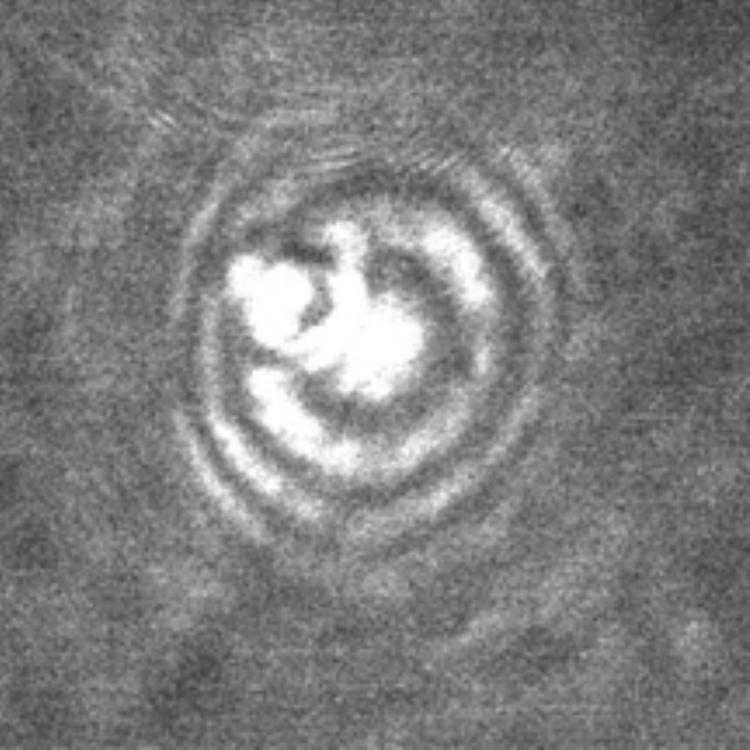}}
    \caption[250ringbead]{A single trapped 1.1 $\mu$m particle trapped in the secondary ring for a 250 $\mu$m thickness cover slip in Configuration 1.}
\label{250ringbead}
\end{figure}

This observation clearly contradicts the normal model of optical trapping of dielectric objects by Gaussian beams. At best, simultaneous trapping of multiple particles would have led to clustering of particles at the trap center and not the assembly of particles in an axis-symmetric ring. In fact, we observe such clustering with 160 $\mu$m slides in both Configuration 1 and 2 as shown in Fig.~\ref{clumping}. 
An obvious step is to image the trapping field for both 250 and 160 $\mu$m cover slips. It is important to note, though, that the imaged field will be a superposition of the fields at all planes in the sample, containing both scattered fields from the micro-sphere, and reflected fields from the different chamber surfaces. Fig.~\ref{160ring} and \ref{250ring} show images recorded by the CCD camera at the microscope back port for 160 and 250 $\mu$m cover slips respectively. The difference in the two images are obvious - in the 160 $\mu$m case the field looks more like that expected for a focused Gaussian beam with a central intensity maxima and faint Airy rings visible. The situation is different in the 250 $\mu$m case, where the central maxima is surrounded by a distinct secondary maxima. 
In fact, the ring formation is likely due to particles getting trapped in the secondary maxima, as shown in Fig.~\ref{250ringbead}, where we have a single 1.1 $\mu$m diameter bead trapped in that region. The question now arises as to why the intensity distribution is so different for the different cover slips. While there are reports of weak trapping of particles in Airy fringes produced by the optics of the trapping system \cite{Mac01}, we do not observe such effects in the 160 $\mu$m cover slips. On the other hand, the ring formation of particles for 250 $\mu$m is extremely stable and does not show any dependence on the power of the trapping laser. Also, the diameter of a ring can be controlled by changing the focus of the objective lens, which means we do have axial trapping. Changing the distance of the objective lens from the sample moves the focus up or down and this changes the diameter of the ring. The rings are formed about 6-10 $\mu$m below the top glass slide, or around 20-24 $\mu$m inside the sample chamber for a sample thickness of 30 $\mu$m. This calibration was performed by the $z$-axis vernier of the microscope (least count 1 $\mu$m). It is also important to note that rings are not observed for sample thicknesses of more than 40 $\mu$m in Configuration 1 and never in Configuration 2. We do not observe any ring formation without the presence of a top surface, i.e. in the absence of a glass slide on top of  the cover slip.

To find the stiffness of our optical trap, we performed measurements of the corner frequency of a single bead trapped in the central maxima, as well as that of a bead trapped in the ring.  A standard measurement of corner frequency can be performed by the power spectrum method \cite{berg04}. It is well known that a trapped probe executing Brownian motion obeys a simplified Langevin equation so that the power spectrum of the probe motion is a Lorentzian. Now, a position sensitive detector or a quadrant photodiode is used to record the displacement of the probe using a detection laser (in our case a laser at 780 nm whose power was kept low enough so as to not affect the trapping) so that the power spectrum can then be obtained. The corner frequency is also a direct measure of the trapping force \cite{Neu04}, and we obtain values of around 165 Hz for a single trapped bead at the beam focus for a glass cover slip, and a maximum of around 25 Hz for a bead trapped in a ring for a polymer cover slip.

From the findings above, it is apparent that we need to understand the reason why the electric field in the sample chamber in Configuration 1 is so different for 160 and 250 $\mu$m cover  slips.  It is thus obvious that a theoretical model of our system is needed. 

\section{\label{analys}THEORETCIAL MODELING}


It is clear that the total field in our sample chamber will be a superposition of forward and backward propagating waves with respect to the sample chamber. What really complicates the problem is the radial variation of the beam that arises from its transverse distribution. 
While there exists literature on the axial distribution of the light field \cite{Mgu92, AAR07}, the radial distribution has not been adequately investigated in optical traps. We use a similar model as in Refs.~\onlinecite{Mgu92, AAR07}, but extend it in the radial direction to find the field distribution inside the sample chamber under various circumstances. Accordingly, we use the well known Angular Spectrum Method (also referred to as vectorial Debye diffraction theory or Debye integral) \cite{Born} to calculate the electric field distribution for high numerical operture optics in stratified media without the use of the paraxial approximation. The approach basically consists of using proper boundary conditions that are necessary to propagate electric fields across multiple interfaces between layered media incorporating the effects of high numerical aperture lenses on refracted fields. We finally use the vectorial electric field distribution integrals for the forward and backward traveling fields inside a particular media with the appropriate Fresnel coefficients for multiple interfaces.

\begin{figure}[ht]
 \centering{\includegraphics[scale=0.45]{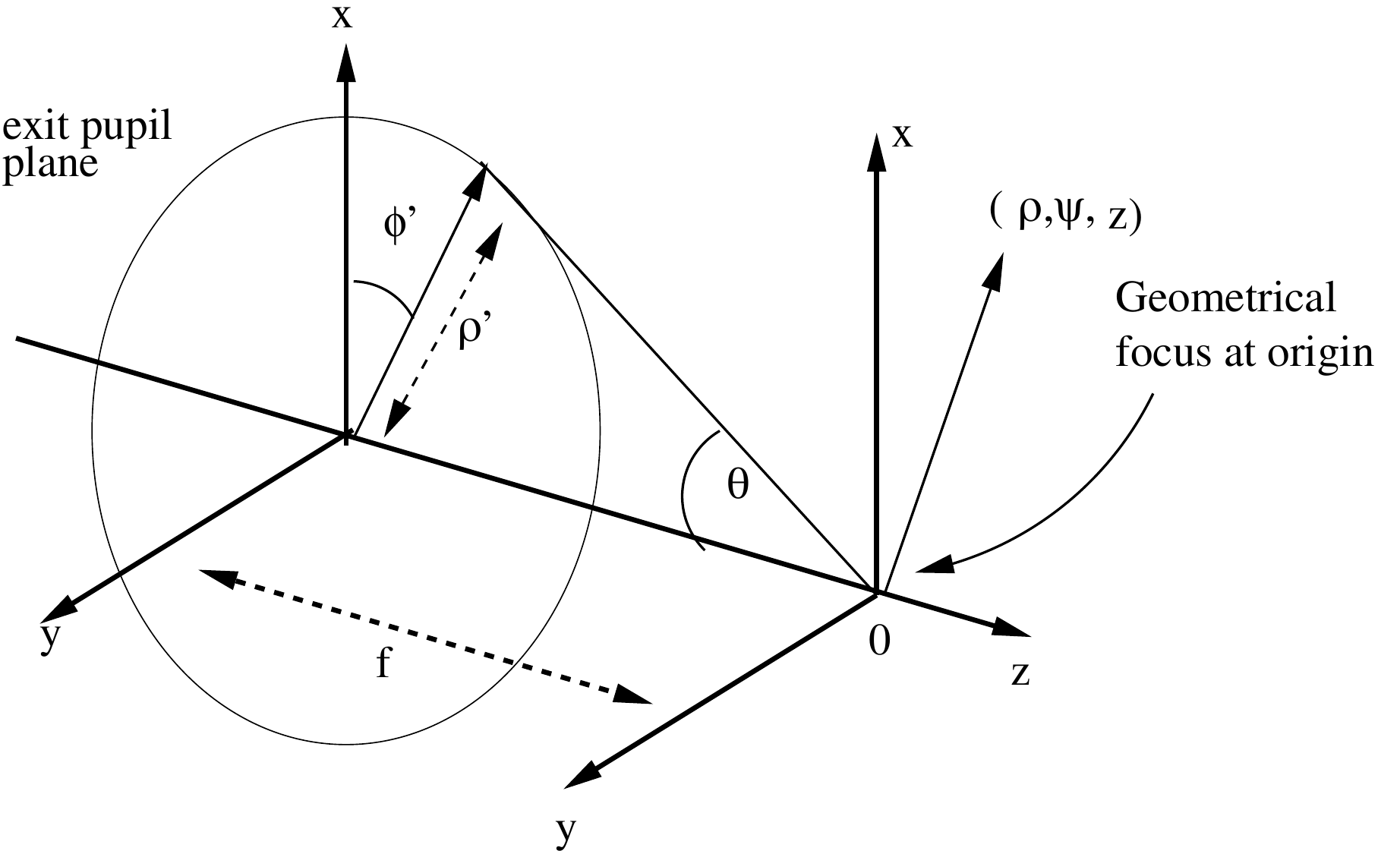}}
    \caption[csystem]{Coordinate system for the problem}
\label{coord}
\end{figure}

The field at the focus of an aplanatic lens (which, for our case, would be the microscope objective) is given by the angular spectrum integral \cite{Born}
\begin{eqnarray}\label{polarint}
\vec{E}(\rho,\psi,z)=  i\frac{kfe^{-ikf}}{2\pi}\int_0^{\theta_{max}}\int_0^{2\pi}
\vec{E}_\infty(\theta,\phi)e^{ikz\cos\theta}e^{ik\rho\sin\theta\cos(\phi-\psi)}
sin(\theta)\>d\theta d\phi
\end{eqnarray}
where $r$ is set to $f$ the focal length of the lens, since the integration is over the spherical wavefront of radius $f$. The limit for the $\theta$ integral is set by the numerical aperture of the microscope objective. The coordinate system used is shown in Fig.~\ref{coord}. Note that the electric field is calculated in cylindrical coordinates, while the $k$ vectors are represented in spherical polar coordinates. The choice of cylindrical coordinates for the electric field makes it convenient to track the polarization of the light beam at the output of a high numerical aperture objective, where it is completely modified from the incident polarization \cite{Roh05}. We have assumed the incident polarization to be in the $x$ direction in cartesian coordinates. The final polarization $E_\infty(\rho, \phi)$ is related to the incident polarization 
$E_{inc}(\rho',\phi')$ by 
\begin{equation}
\vec{E}_{\infty}(\rho',\phi')=E_{inc}(\rho',\phi')\left(\cos\phi'\hat{\theta}-\sin\phi'\hat{\phi}'\right)
\end{equation}
Using the cartesian form of the unit vectors $\hat{\theta}$ and $\hat{\phi}$, we obtain
\begin{eqnarray}
\hat{\theta}=\cos\theta\cos\phi\hat{i}+\cos\theta\sin\phi\hat{j}+\sin\theta\hat{k}\nonumber \\
 \hat{\phi}=\hat{\phi}'=-\sin\phi\hat{i}+\cos\phi\hat{j}
\end{eqnarray}
so that the final  polarization can be written as\\
\begin{eqnarray}
\vec{E}_\infty(\theta,\phi)=\dfrac{E_{inc}(\theta,\phi)}{2} \left [
\begin{array}{c}
\{(1+\cos\theta)-(1-\cos\theta)\cos(2\phi)\}\hat{i}\\
\{-(1-\cos\theta)\sin(2\phi)\}\hat{j}\\
\{-2\cos\phi\sin\theta\}\hat{k}
\end{array}
\right ] \sqrt{\cos\theta}
\end{eqnarray}
\\ where the $\sqrt{\cos\theta}$ term came from the apodization function for an aplanatic lens \cite{Mgu92}. Now, in general, $\vec{E}_{inc}$ can have a phase curvature. However, to keep things simple, we assume
that it hits the entrance pupil of the lens with a planar phase front perpendicular to the optical axis. Also if the beam is a fundamental Gaussian, then the intensity distribution is independent of $\phi$, and
equation \ref{polarint} can be written as
\begin{displaymath}
 \vec{E}(\rho,\psi,z)=\frac{ikf}{2}e^{-ikf}
\left[
\begin{array}{c}
 \{I_0+I_2 cos(2\psi)\}\hat{i}\\
\{I_2 sin(2\psi)\}\hat{j}\\
\{i2I_1 cos(\psi)\}\hat{k}\\
\end{array}
\right ]
\end{displaymath}
where,
\begin{eqnarray}
I_0 = \int_0^{\theta_{max}}E_{inc}(\theta)\sqrt{\cos\theta}(1+\cos\theta)J_0(k\rho\sin\theta)e^{ikz\cos\theta}
\sin\theta\>d\theta 
\end{eqnarray}
\begin{equation}
I_1=\int_0^{\theta_{max}}E_{inc}(\theta)\sqrt{\cos\theta}J_1(k\rho\sin\theta)e^{ikz\cos\theta}\sin^2\theta\>d\theta
\end{equation}
\begin{eqnarray}
I_2=\int_0^{\theta_{max}}E_{inc}(\theta)\sqrt{\cos\theta}(1-\cos\theta)J_2(k\rho\sin\theta)e^{ikz\cos\theta}\sin\theta\>d\theta 
\end{eqnarray}
where the $\phi$ integrals have been carried out and related to Bessel functions $J_n$.

\begin{figure}[ht]
 \centering{\includegraphics[scale=0.45]{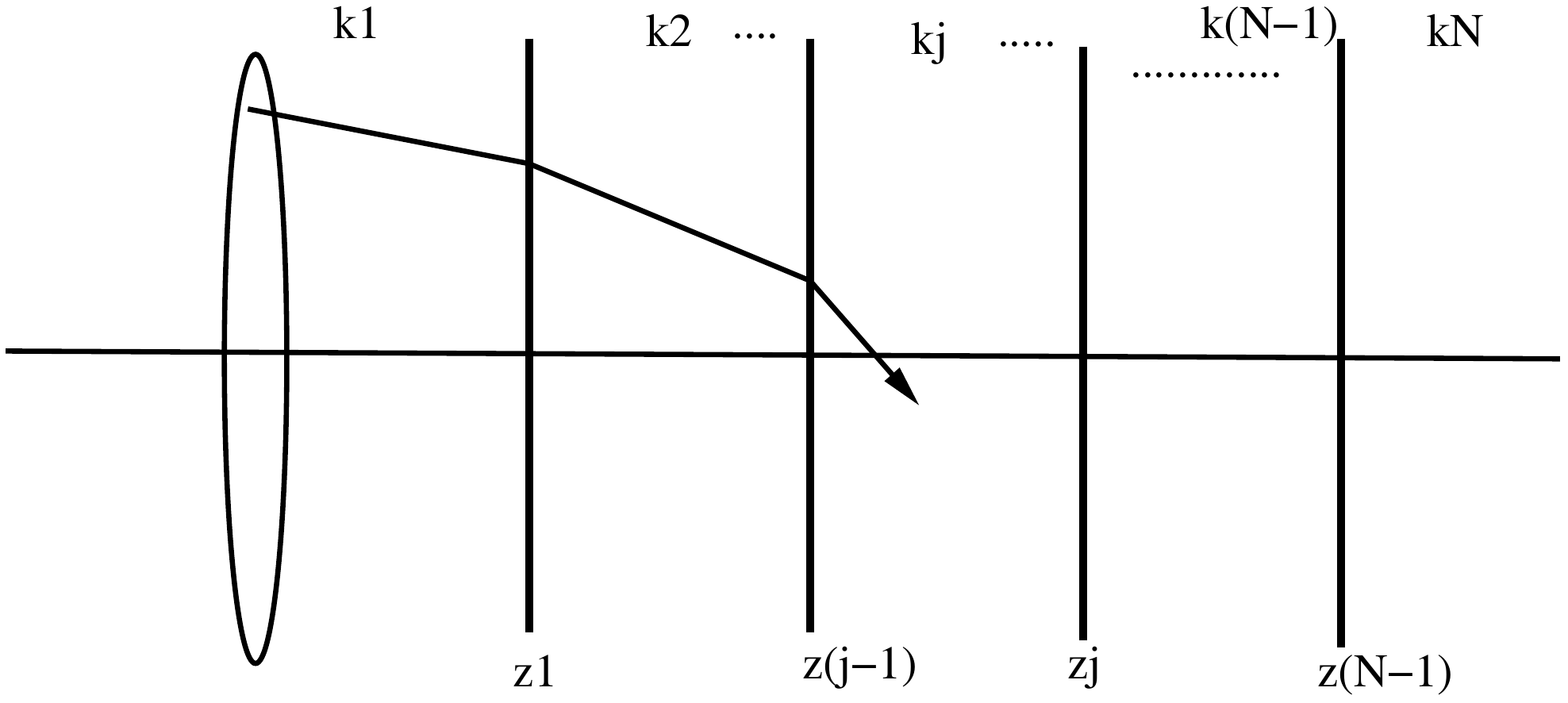}}
    \caption[nlayerlens]{Aplanatic lens focusing inside the $j$ th medium in a $N$ layered stratified media}
\label{nlayerlens}
\end{figure}

Thus, with the vector field equations at the focus of an aplanatic lens known, we now consider the aplanatic lens placed in front of a multi-layered media, with the focused light beam propagating through n different dielectric interfaces as shown in Fig.~\ref{nlayerlens}.
The integrals for finding the forward and backward traveling fields inside the $j$ th media can then be written down directly
 \begin{displaymath}
 \vec{E_t}(\rho,\psi,z)=\frac{ikf}{2}e^{-ikf}
\left[
\begin{array}{c}
\{I^t_0+I^t_2cos(2\psi)\}\hat{i}\\
\{I^t_2\>sin(2\psi)\}\hat{j}\\
\{i2\>I^t_1 cos(\psi)\}\hat{k}\\
\end{array}
\right ]
\end{displaymath}
where,
\begin{eqnarray}\label{transmisseqn}
I^t_0=\int_0\limits^{min(\theta_{max},\theta_c)}E_{inc}(\theta)
\sqrt{\cos\theta}(T^{(1,j)}_s+T^{(1,j)}_p\cos\theta_j)J_0(k_1\rho\sin\theta)e^{ik_jz\cos\theta_j}sin(\theta)\>d\theta \nonumber
\end{eqnarray}
\begin{eqnarray}
I^t_1=\int_0\limits^{min(\theta_{max},\theta_c)}E_{inc}(\theta)
\sqrt{\cos\theta}T^{(1,j)}_p\sin\theta_j
J_1(k_1\rho\sin\theta)e^{ik_jz\cos\theta_j}\sin\theta\>d\theta \nonumber
\end{eqnarray}
\begin{eqnarray}
I^t_2=\int_0\limits^{min(\theta_{max},\theta_c)}E_{inc}
(\theta)\sqrt{\cos\theta}(T^{(1,j)}_s-T^{(1,j)}_p\cos\theta_j) J_2(k_1\rho\sin\theta)e^{ik_jz\cos\theta_j}\sin\theta\>d\theta
\end{eqnarray}
where we use the Fresnel coefficients $T_s$ and $T_p$ for multiple interfaces. These are related to the Fresnel coefficients $t_i$ and $r_i (i = s, p$ polarizations) for a single interface by
\begin{eqnarray}
T^{(n-1,n+1)}_i&=&\sum\limits^{j-1}_{n=1}\left (\dfrac{t^{(n-1,n)}_it^{(n,n+1)}_ie^{i\beta}}
{1+r^{(n-1,n)}_ir^{(n,n+1)}_ie^{i2\beta}}\right )\nonumber\\
&&e^{i(z_{n-1}k_{n-1}\cos\theta_{n-1}-z_nk_{n+1}\cos\theta_{n+1})}\nonumber\\
R^{(n-1,n+1)}_i&=&\sum\limits^{j-1}_{n=1}\left (\dfrac{r^{(n-1,n)}_i+r^{(n,n+1)}_ie^{i2\beta}}
{1+r^{(n-1,n)}_ir^{(n,n+1)}_ie^{i2\beta}}\right)\nonumber\\
&&e^{i2z_{n-1}k_{n-1}
\cos\theta_{n-1}}
\end{eqnarray}
 where,
\begin{equation}
 \beta=(z_n-z_{n-1})k_n\cos\theta_n
\end{equation}.
It is interesting to note that an aberration term  is also present in Eq.~\ref{transmisseqn} hidden inside the transmission coefficients, and is given by
\begin{equation}
\Phi(\theta)= \sum\limits^{j-1}_{n=1}z_n(k_n\cos\theta_n-k_{n+1}\cos\theta_{n+1})
\end{equation}
where $z_n$ is the location of the $nth$ interface. This is basically the well-known spherical aberration term that appears due to refractive index mismatch at different dielectric interfaces.
The reflected field integral inside the $jth$ medium is also calculated similarly and is given by
\begin{displaymath}
 \vec{E}(\rho,\psi,z)=\frac{ikf}{2}e^{-ikf}
\left[
\begin{array}{c}
 \{I^r_0+I^r_2 cos(2\psi)\}\hat{i}\\
\{I^r_2 sin(2\psi)\}\hat{j}\\
\{-i2\>I^r_1 cos(\psi)\}\hat{k}\\
\end{array}
\right ]
\end{displaymath}
where once again we have
\begin{eqnarray}
I^r_0=\int_0\limits^{min(\theta_{max},\theta_c)}E_{inc}(\theta)\sqrt{\cos\theta}
(R^{(1,j)}_s-R^{(1,j)}_p\cos\theta_j)J_0(k_1\rho\sin\theta)e^{-ik_jz\cos\theta_j}\sin\theta\>d\theta \nonumber
\label{finalint}
\end{eqnarray}
\begin{eqnarray}
I^r_1=\int_0\limits^{min(\theta_{max},\theta_c)}E_{inc}(\theta)\sqrt{\cos\theta}
R^{(1,j)}_p\sin\theta_k J_1(k_1\rho\sin\theta)e^{-ik_jz\cos\theta_j}\sin\theta_1\>d\theta \nonumber
\end{eqnarray}
\begin{eqnarray}
I^r_2=\int_0^{min(\theta_{max},\theta_c)}E_{inc}(\theta)\sqrt{\cos\theta}
(R^{(1,j)}_s+R^{(1,j)}_p\cos\theta_j) J_2(k_1\rho\sin\theta)e^{-ik_jz \cos\theta_j}\sin\theta\>d\theta
\end{eqnarray}
Note that in all subsequent simulations we have used n = 3. 

\section{\label{resdis}RESULTS AND DISCUSSIONS}
\subsection{\label{simulres}SIMULATIONS}
A computer simulation was developed using Eqns.~\ref{transmisseqn} and ~\ref{finalint} to determine the electric field inside our sample chamber. As per specifications, the Zeiss objective focal length was taken to be 1.8 mm or 1800 $\mu$m, and the numerical aperture (NA) 1.41. The wavelength of light was 1.064 $\mu$m, while the velocity of light in vacuum was considered to be 1. The beam waist radius after the objective was assumed to be 0.5 $\mu$m, which we obtained by measuring the beam waist after our trapping objective. This is consistent with the minimum obtainable theoretical resolution corresponding to $\dfrac{\lambda}{2 NA}$, $\lambda$ and NA for our experiment being 1064 nm and 1.41 respectively. We also considered only propagating light waves and not the evanescent components. This is consistent with our experimental observation of not obtaining any ring formation near the cover slip, a fact that rules out any evanescent coupling with the microspheres.  

The model was verified for the field intensity distribution around the focus of a high numerical aperture lens and produced the well known Airy fringes at the beam focus. The field distribution also showed a slight asymmetry or elongation around the direction of polarization of the light beam as expected. However, only Airy fringes are not sufficient to trap particles away from the center since the power in these fringes is very low compared to that in the central maxima. Next, we considered an aplanatic lens focusing into a single interface (glass/water). Again as expected, we saw signatures of spherical aberration, with the focal spot shifting from the geometric focus, and more importantly, some power being redistributed in the form of side lobes in the beam. Note that for both these cases, only Eqns.~\ref{transmisseqn} were used since there was no reflected component of the field.

Finally, we considered three interfaces akin to our experimental sample chamber in Configuration 1. After the microscope objective, the three interfaces are: 1) index-matching microscope objective oil and cover slip (English glass or polymer), 2) cover slip and sample (water), 3) water and top slide (glass). To determine the effect of the 160 and 250 $\mu$m cover slips in the radial distribution of the field for sample thicknesses used in our experiment, we did not consider a third interface.  The thickness of the sample (water layer) for all simulations was 30 $\mu$m, the same as the experimental value used. The geometric focus was placed 15$\mu$m inside the water layer in each case, the distance being measured from the cover slip-water interface. Only the results for the transmitted field are given. For Fig.~\ref{160focaldist}, \ref{160focalplots} and~\ref{250focalplots}, only Eqn.~\ref{transmisseqn} was used, while for Fig.~\ref{axialdist1}, the field inside the chamber was calculated by a superposition of Eqns.~\ref{transmisseqn} and ~\ref{finalint}.

\subsection{\label{anal}ANALYSIS}
\begin{figure}[h]
  \centering
\subfloat[]{\includegraphics[width=0.55\textwidth]{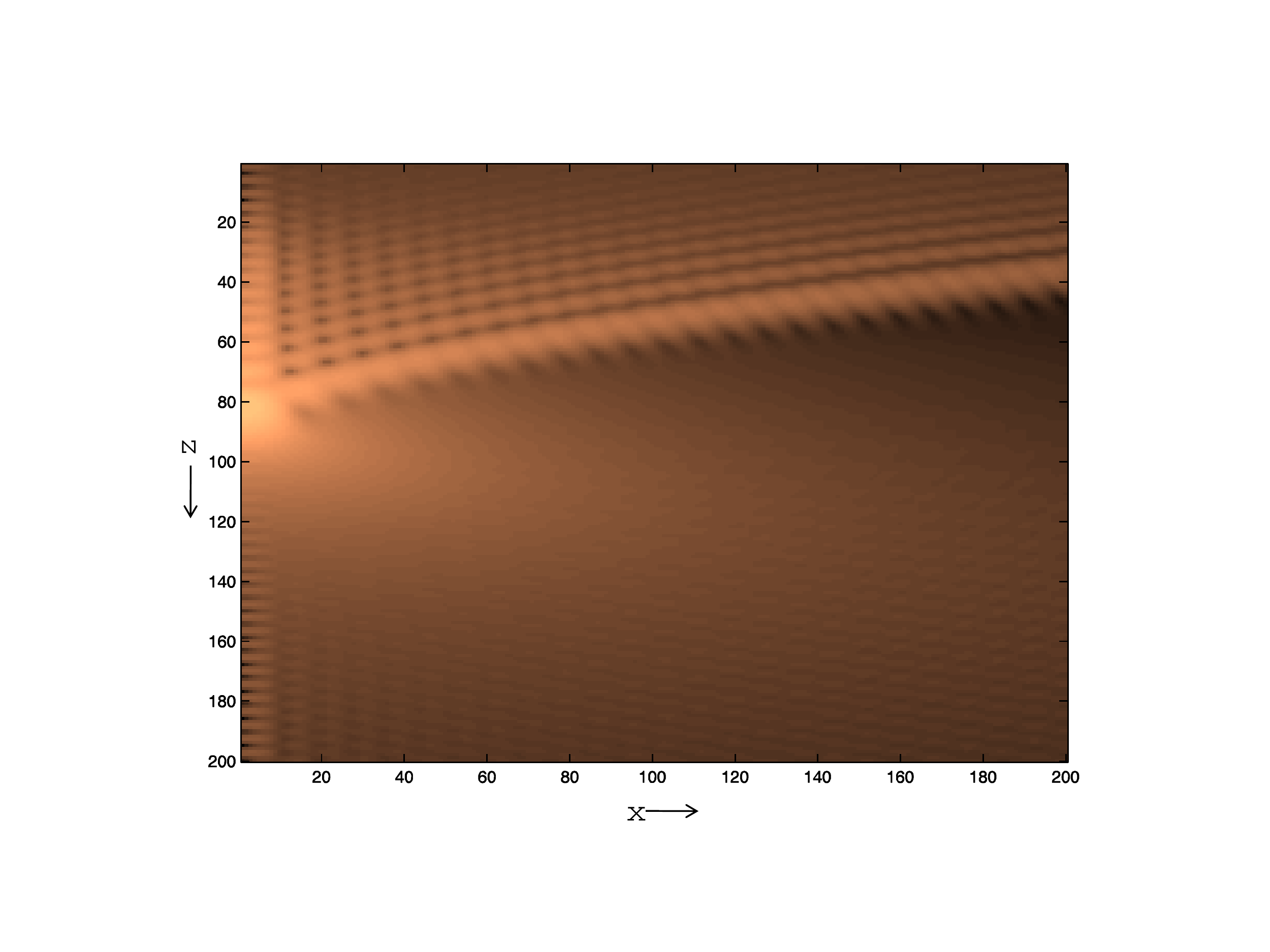}\label{xza3}}
  \subfloat[]
{\includegraphics[width=0.56\textwidth]{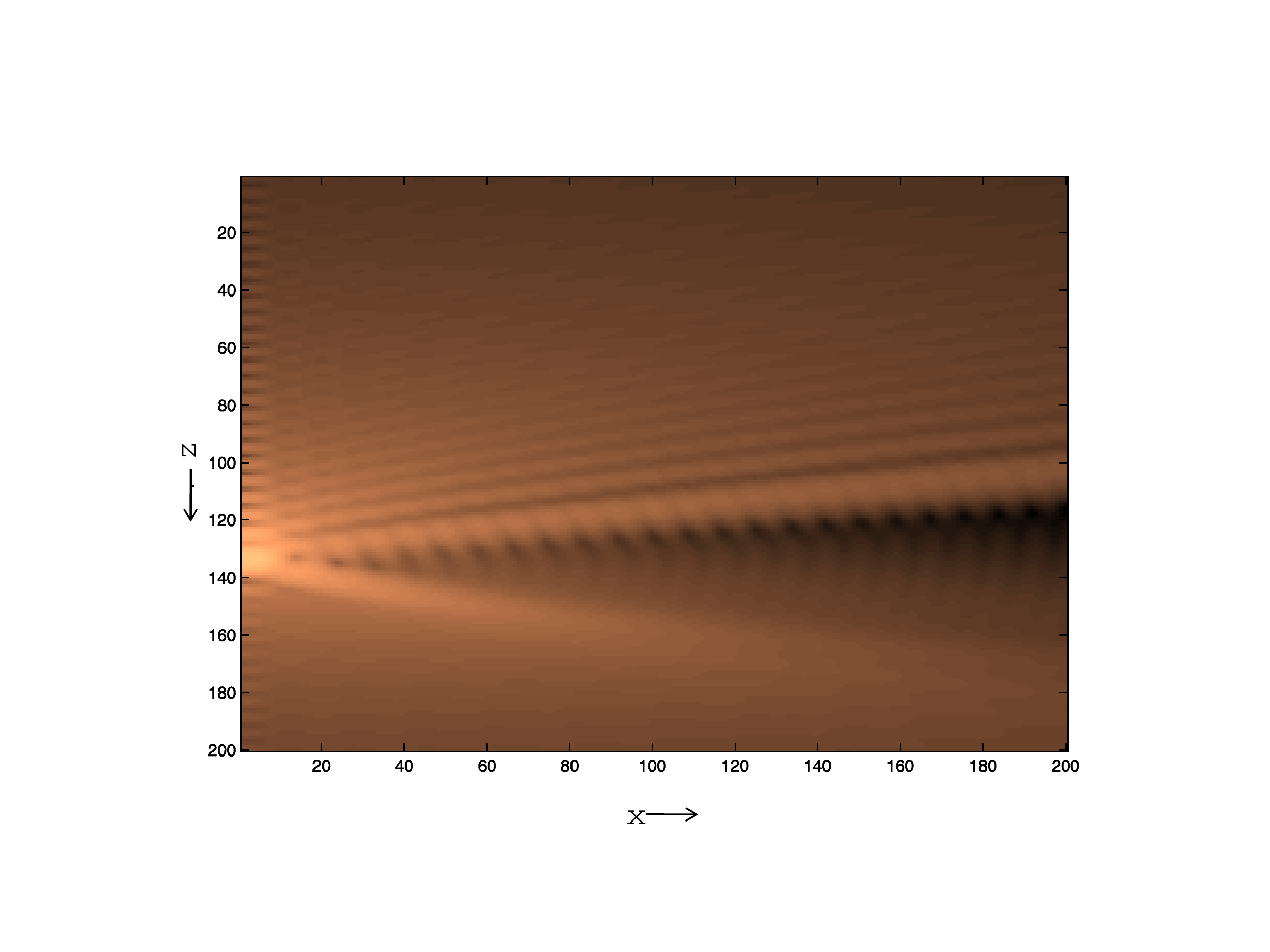}\label{xza250}}
  \caption{XZ intensity plot for (a) 160 $\mu$m cover slip, and (b) 250 $\mu$m cover slip ($0<x<10~\mu$m,$-15<z<15~\mu$m.) in Configuration 1. The trapping beam propagates along the z-direction. Both plots show distortions of the focal spots with power being redistributed into side lobes. In (a), the focal spot has shifted in the negative z-direction and side lobes arise in the backward direction with respect to beam propagation, while in (b), the focal spot has shifted in the positive z-direction and side lobes arise in the forward direction with respect to beam propagation.}
  \label{160focaldist}
\end{figure}
Fig.~\ref{160focaldist} shows the results for the field profile in the $xz$ plane obtained for 160 and 250 $\mu$m cover slips.  Both plots show considerable aberrations, which results in the distortion of the focal spot and distributes more power to the side lobes. For 160 $\mu$m cover slips, as shown in Fig.~\ref{xza3}, the focal spot has shifted towards the negative z-direction, i.e. towards the cover slip, and is now at around 13 $\mu$m (a change of -2 $\mu$m) inside the water layer of the sample chamber. The plot clearly shows the side lobes that arise in the backward direction due to excess aberration. In contrast, for 250 $\mu$m cover slips as shown in Fig.~\ref{xza250}, the focus has shifted in a direction opposite to that obtained in the former case, i.e. in the positive z-direction, and is now at 20 $\mu$m (a change of 5 $\mu$m). The plot clearly shows the side lobes that arise in this case in the forward direction due to excess aberration. 

\begin{figure}[h]
  \centering
{\includegraphics[scale=0.55]{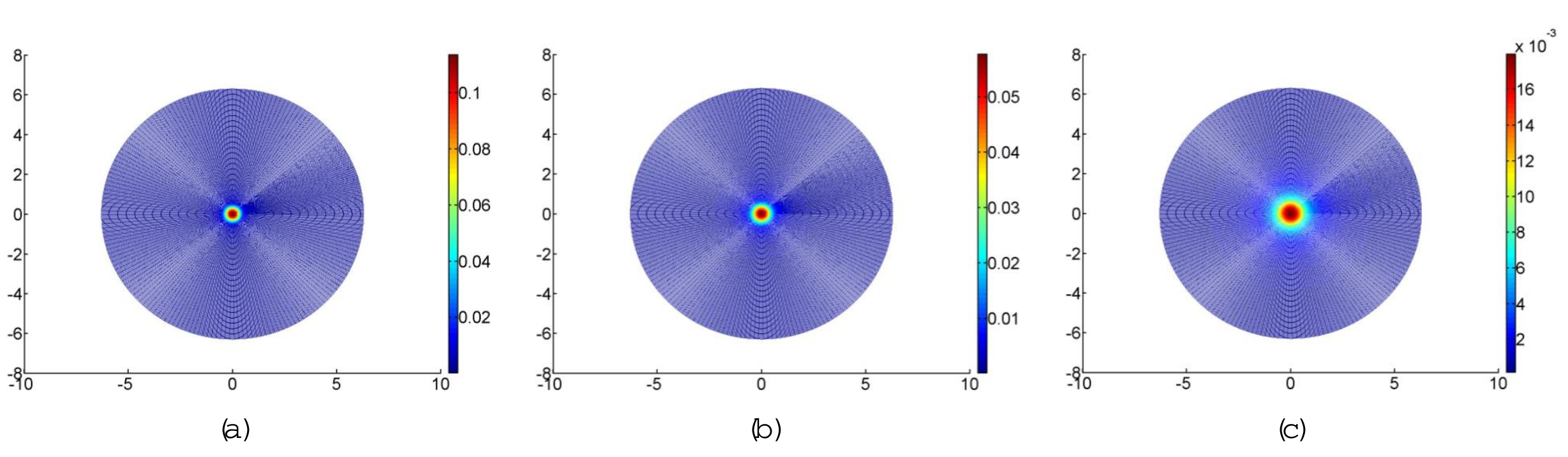}}
  \caption{Radial intensity plot inside the sample chamber at a plane (a) $1\mu$m, (b) $2\mu$m, and (c) $3\mu$m away from focus for a 160 $\mu$m cover slip. The sample (water with polystyrene beads) thickness is 30 $\mu$m. The focus is situated at around 13 $\mu$m inside the cover slip. As is clear, the maximum field intensity is concentrated in the central region for all three plots. The x-axis is in microns. }
 \label{160focalplots}
\end{figure}
\begin{figure}[!h]
  \centering
{\includegraphics[scale=0.55]{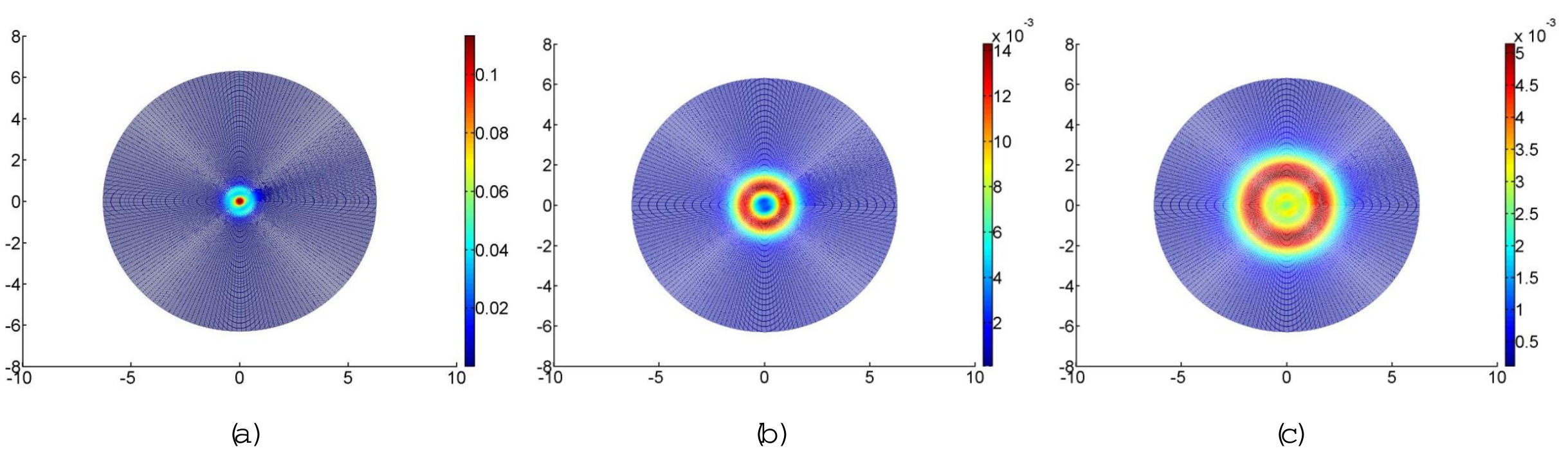}}
  \caption{Radial intensity plot inside the sample chamber at a plane (a) $1\mu$m, (b) $2\mu$m, and (c) $3\mu$m away from focus for a 250 $\mu$m cover slip. The sample (water with polystyrene beads) thickness is 30 $\mu$m. The focus is situated at around 20 $\mu$m inside the cover slip. In plots (b) and (c), the intensity is no longer highest in the center, but in a ring of diameter of around 3 $\mu$m for plot (b) and 5 $\mu$m for plot (c). The x-axis is in microns. }
 \label{250focalplots}
\end{figure}
A crucial difference between two types of cover slips is that for the 160 $\mu$m case, the intensity is almost completely concentrated in the center, with very little intensity in the side lobes. This has been tested out by taking 3d plots of the electric field inside the sample chamber at various distances from the actual focus for both 160 and 250 $\mu$m cover slips. Fig.~\ref{160focalplots} shows the field distribution at distances 1, 2 and 3 $\mu$m away from the focus for the 160 $\mu$m case. As is clear, the intensity is highest in the beam center, with the diameter of the beam expanding as it propagates beyond the focus. However, for 250 $\mu$m cover slips, the situation is very different as is shown in Fig.~\ref{250focalplots}. Again, we plot at 1, 2, and 3 $\mu$ distances from the focus, and observe that for the 2 and 3 $\mu$m cases (Fig.~\ref{250focalplots}(b) and \ref{250focalplots}(c)), the maximum intensity is not in the center, but actually in the secondary maximum away from the center. The diameter of the secondary maxima is also around 3 $\mu$m in Fig.~\ref{250focalplots}(b) and 5 $\mu$m in Fig.~\ref{250focalplots}(c), which matches the ring diameters we obtain experimentally. This gives us confidence about the authenticity of our modeling and our understanding of the phenomenon behind the ring formation.

However, the radial fringes are still not enough to support axial trapping, as is evident from our experimental findings where we do not obtain ring structures with just the sample on a cover slip. This tells us that there is still a need to accentuate axial trapping for achieving stable trapping in a ring. The enhanced axial trapping is provided by the top slide. Enhanced axial trapping due to the presence of a reflective surface has been studied in Ref.~[\onlinecite{Zem01}]. That study dealt with a similar set-up such as ours with a trapping chamber consisting of the sample sandwiched between cover slip and a top slide. It was shown that the top and bottom surface form a standing wave cavity with interference fringes formed by superposition of the transmitted trapping light and that reflected from the top slide of the sample chamber causing alternate regions of stable axial trapping near the slide. The authors calculated the nodes of the trapping force developed near the top slide and experimentally showed trapped particles hopping between different stable equilibrium axial positions. However, they did not consider the radial component of the trapping field in their theoretical calculations, and did not experimentally observe any confinement away from the trap center. The enhanced spherical aberration in our system provides the radial confinement, while the top slide stabilizes the trap axially. To investigate this by a simulation, we consider a four layered media now, with the fourth media representing the glass slide having a refractive index of 1.516.
\begin{figure*}[h]
  \centering
{\includegraphics[scale=0.25]{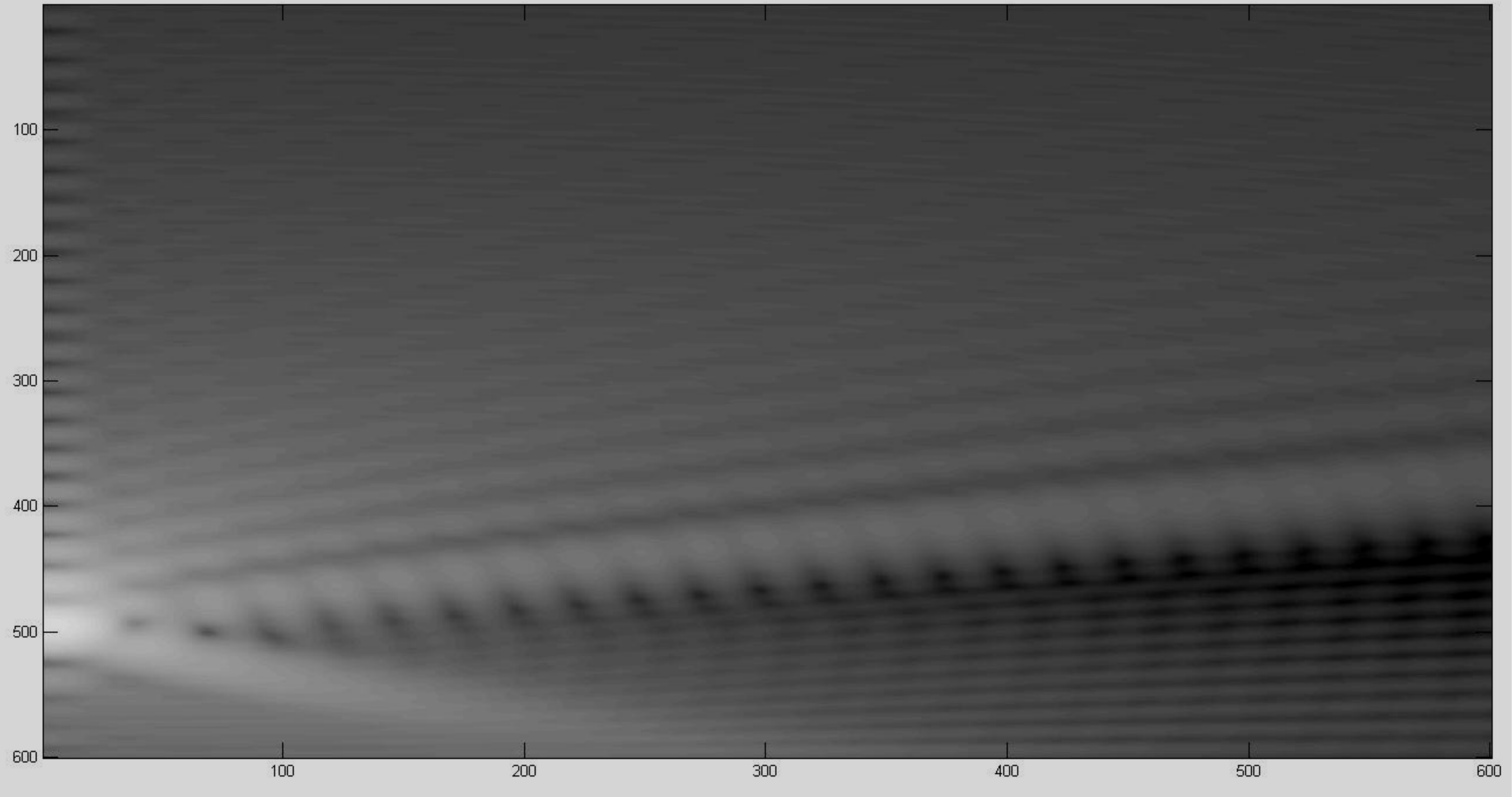}}
  \caption{Axial fringes produced due to reflections from top slide ($0<x<10~\mu$m,$-15m<z<15~\mu$m).}
  \label{axialdist1}
\end{figure*}
The top slide contributes to reflected waves that causes additional axial fringes to develop within the electric field distribution as shown in the $xz$ plot in Fig.~\ref{axialdist1}. The location and separation of the fringes depends on the thickness of the water layer and also on the position of the focus with respect to the top glass slide - water interface. The figure was generated with focal spot 6 $\mu$m below the glass slide water interface.

It is also clear why the rings are observed only when the beam focus is close to the top glass slide. As is apparent from our simulations shown in Fig.~\ref{axialdist1}, as well as in the results given in Ref.~[\onlinecite{Zem01}], the axial fringes are located near the top slide only and die away very fast as one goes farther into the sample solution. This is intuitively understandable considering the fact that we are working with very fast diverging Gaussian beams in this case, and a constructive superposition could be achieved only when the incident beam and reflective surface are very close. This also explains why we did not observe the patterns at sample thicknesses of more than around 30 $\mu$m. For greater distances, we could not get the beam focus close enough to the top slide to facilitate axial trapping, as a result of which the ring structures were not axially stable.

\section{\label{concl}Conclusions}
The self-assembly we observe could be used for several interesting applications, the most obvious being the study of optical binding. This is the phenomenon where the equilibirum position of a trapped microparticle is modified due to the presence of a second trapped particle in its vicinity \cite{Dhol10}. The electric field due to one microparticle, which is actually an induced dipole, alters the electric field perceived by the other, and therefore alters its equilibrium position. An important requirement to see the effects of optical binding in an optical trap is that the trap stiffness should not be very high, since the binding forces are only in the scale of 0.1 - 1 pN \cite{Brz10}. A strong optical trap would not allow an experimental determination of such a weak interaction force as Ref.~[\onlinecite{Brz10}] points out. As we mentioned in Section ~\ref{expmt}, the corner frequency we obtain is around 25 Hz for polystyrene beads trapped in the ring, that corresponds to a trap stiffness of around 1.3pN/$\mu$m \cite{berg04} - which should be ideal to study optical binding experimentally. One could in principle, continuously monitor the corner frequency of a trapped microparticle and study how it gets modified due to the entry of other particles into the ring. The measurement could be performed by a high speed camera or even by a standard position detector such as a quadrant photodiode. Thus, direct measurements of optical binding could be facilitated by such structures. 

The ring pattern can be of use in biological applications as well, with the possibility of studying controllable cell to cell interactions, such as that between cancer cells and Natural Killer cells \cite{Old83}. Also, the possibility of applying differential force in the center and the sides of a single trapped cell could facilitate measurements of cell elasticity, where the cell is stretched and then unstretched by switching off the trap momentarily (which may be accomplished by modulating the trap by an acousto-optic modulator). 

Future work would revolve around the applications mentioned above. However, one of the immediate things we are working on is the calculation of the forces in the axial and radial directions by using the Generalized Lorentz Mie theory \cite{Gou88}. While the intensity calculations shown in this paper give us a clear understanding of the phenomenon we report, a detailed force calculation would also give more quantitative understanding and also identify the stable positions where particles could be trapped in the axial and radial directions in the trapping chamber in an optical trap.

In conclusion, we have shown spontaneous assembly of 1.1 $\mu$m particles in a ring formation for single beam optical tweezers without the use of spatially manicured optical beams or holographic patterns coupled into the optical trap. The patterns are observed in a standard trapping chamber consisting of a cover slip stuck to a top slide with the sample consisting of 1.1 $\mu$m diameter polystyrene beads in water solution sandwiched between them. The patterns are seen only when we use polymer cover slips that are thicker and have higher refractive index than glass cover slips typically used for optical trapping. This is due to the fact that spherical aberration comes into play because of the enhanced thickness of the polymer cover slips and the mismatch of the refractive indices between the immersion oil, cover slip and water in which the beads were suspended. By a simulation of the electric field inside our sample chamber, we demonstrate that the net spherical aberration is able to elongate the beam focus axially so that the intensity in the central region is reduced and more power is diverted into side lobes that are formed radially, thus increasing the chances of ring formation. The axial trapping is provided by top glass slide. The reflected light from the glass slide produces axial interference fringes on the radial fringes existing already, which in turn creates pockets of high intensity such that the trapping of the micro-spheres gets axially stabilized within the rings. We believe that this kind of assembly formation due to only optical effects is quite novel and has not been observed before. We envisage several applications for this assembly including optical binding, and the study of biological interactions.
\section{Acknowledgements}
This work was supported by the Indian Institute of Science Education and Research, Kolkata, an autonomous research and teaching institute funded by the Ministry of Human Resource Development, Govt. of India. The authors would also like to acknowledge Dr. Nirmalya Ghosh of IISER-Kolkata for useful technical discussions and comments about the manuscript, and Dr. Kaushik Biswas and Dr. Basudeb Karmakar of CGCRI Kolkata, for help in measuring the refractive indices of the glass and polymer cover slips. 

\nocite{*}


\begin{thebibliography}{24}%
\makeatletter
\providecommand \@ifxundefined [1]{%
 \@ifx{#1\undefined}
}%
\providecommand \@ifnum [1]{%
 \ifnum #1\expandafter \@firstoftwo
 \else \expandafter \@secondoftwo
 \fi
}%
\providecommand \@ifx [1]{%
 \ifx #1\expandafter \@firstoftwo
 \else \expandafter \@secondoftwo
 \fi
}%
\providecommand \natexlab [1]{#1}%
\providecommand \enquote  [1]{``#1''}%
\providecommand \bibnamefont  [1]{#1}%
\providecommand \bibfnamefont [1]{#1}%
\providecommand \citenamefont [1]{#1}%
\providecommand \href@noop [0]{\@secondoftwo}%
\providecommand \href [0]{\begingroup \@sanitize@url \@href}%
\providecommand \@href[1]{\@@startlink{#1}\@@href}%
\providecommand \@@href[1]{\endgroup#1\@@endlink}%
\providecommand \@sanitize@url [0]{\catcode `\\12\catcode `\$12\catcode
  `\&12\catcode `\#12\catcode `\^12\catcode `\_12\catcode `\%12\relax}%
\providecommand \@@startlink[1]{}%
\providecommand \@@endlink[0]{}%
\providecommand \url  [0]{\begingroup\@sanitize@url \@url }%
\providecommand \@url [1]{\endgroup\@href {#1}{\urlprefix }}%
\providecommand \urlprefix  [0]{URL }%
\providecommand \Eprint [0]{\href }%
\@ifxundefined \urlstyle {%
  \providecommand \doi  [0]{\begingroup \@sanitize@url \@doi}%
  \providecommand \@doi [1]{\endgroup \@@startlink {\doibase
  #1}doi:\discretionary {}{}{}#1\@@endlink }%
}{%
  \providecommand \doi  [0]{doi:\discretionary{}{}{}\begingroup
  \urlstyle{rm}\Url }%
}%
\providecommand \doibase [0]{http://dx.doi.org/}%
\providecommand \Doi [0]{\begingroup \@sanitize@url \@Doi }%
\providecommand \@Doi  [1]{\endgroup\@@startlink{\doibase#1}\@@Doi}%
\providecommand \@@Doi [1]{#1\@@endlink}%
\providecommand \selectlanguage [0]{\@gobble}%
\providecommand \bibinfo  [0]{\@secondoftwo}%
\providecommand \bibfield  [0]{\@secondoftwo}%
\providecommand \translation [1]{[#1]}%
\providecommand \BibitemOpen [0]{}%
\providecommand \bibitemStop [0]{}%
\providecommand \bibitemNoStop [0]{.\EOS\space}%
\providecommand \EOS [0]{\spacefactor3000\relax}%
\providecommand \BibitemShut  [1]{\csname bibitem#1\endcsname}%
\bibitem [{\citenamefont {Ghislain}\ \emph {et~al.}(1994)\citenamefont
  {Ghislain}, \citenamefont {Switz},\ and\ \citenamefont {Webb}}]{Ghi94}%
  \BibitemOpen
  \bibfield  {author} {\bibinfo {author} {\bibfnamefont {L.~P.}\ \bibnamefont
  {Ghislain}}, \bibinfo {author} {\bibfnamefont {N.~A.}\ \bibnamefont {Switz}},
  \ and\ \bibinfo {author} {\bibfnamefont {W.~W.}\ \bibnamefont {Webb}},\
  }\href@noop {} {\bibfield  {journal} {\bibinfo  {journal} {Rev. Sci.
  Instrum.},\ }\textbf {\bibinfo {volume} {65}},\ \bibinfo {pages} {2762}
  (\bibinfo {year} {1994})}\BibitemShut {NoStop}%
\bibitem [{\citenamefont {Volpe}\ and\ \citenamefont {Petrov}(2006)}]{volpe06}%
  \BibitemOpen
  \bibfield  {author} {\bibinfo {author} {\bibfnamefont {G.}~\bibnamefont
  {Volpe}}\ and\ \bibinfo {author} {\bibfnamefont {D.}~\bibnamefont {Petrov}},\
  }\href@noop {} {\bibfield  {journal} {\bibinfo  {journal} {Phys. Rev.
  Lett.},\ }\textbf {\bibinfo {volume} {6975}},\ \bibinfo {pages} {210603}
  (\bibinfo {year} {2006})}\BibitemShut {NoStop}%
\bibitem [{\citenamefont {Clapp}\ \emph {et~al.}(1999)\citenamefont {Clapp},
  \citenamefont {Ruta},\ and\ \citenamefont {Dickinson}}]{clapp99}%
  \BibitemOpen
  \bibfield  {author} {\bibinfo {author} {\bibfnamefont {A.~R.}\ \bibnamefont
  {Clapp}}, \bibinfo {author} {\bibfnamefont {A.~G.}\ \bibnamefont {Ruta}}, \
  and\ \bibinfo {author} {\bibfnamefont {R.~B.}\ \bibnamefont {Dickinson}},\
  }\href@noop {} {\bibfield  {journal} {\bibinfo  {journal} {Rev. Sci.
  Instrum.},\ }\textbf {\bibinfo {volume} {70}},\ \bibinfo {pages} {26}
  (\bibinfo {year} {1999})}\BibitemShut {NoStop}%
\bibitem [{\citenamefont {Svoboda}\ \emph {et~al.}(1993)\citenamefont
  {Svoboda}, \citenamefont {Schmidt}, \citenamefont {Schnapp},\ and\
  \citenamefont {Block}}]{svo94}%
  \BibitemOpen
  \bibfield  {author} {\bibinfo {author} {\bibfnamefont {K.}~\bibnamefont
  {Svoboda}}, \bibinfo {author} {\bibfnamefont {C.~F.}\ \bibnamefont
  {Schmidt}}, \bibinfo {author} {\bibfnamefont {B.~J.}\ \bibnamefont
  {Schnapp}}, \ and\ \bibinfo {author} {\bibfnamefont {S.~M.}\ \bibnamefont
  {Block}},\ }\href@noop {} {\bibfield  {journal} {\bibinfo  {journal}
  {Nature},\ }\textbf {\bibinfo {volume} {365}},\ \bibinfo {pages} {721}
  (\bibinfo {year} {1993})}\BibitemShut {NoStop}%
\bibitem [{\citenamefont {Mehta}\ \emph {et~al.}(1999)\citenamefont {Mehta},
  \citenamefont {Rief}, \citenamefont {Spudich}, \citenamefont {Smith},\ and\
  \citenamefont {Simmons}}]{mehta99}%
  \BibitemOpen
  \bibfield  {author} {\bibinfo {author} {\bibfnamefont {A.~D.}\ \bibnamefont
  {Mehta}}, \bibinfo {author} {\bibfnamefont {M.}~\bibnamefont {Rief}},
  \bibinfo {author} {\bibfnamefont {J.~A.}\ \bibnamefont {Spudich}}, \bibinfo
  {author} {\bibfnamefont {D.~A.}\ \bibnamefont {Smith}}, \ and\ \bibinfo
  {author} {\bibfnamefont {R.~M.}\ \bibnamefont {Simmons}},\ }\href@noop {}
  {\bibfield  {journal} {\bibinfo  {journal} {Science},\ }\textbf {\bibinfo
  {volume} {283}},\ \bibinfo {pages} {1689} (\bibinfo {year}
  {1999})}\BibitemShut {NoStop}%
\bibitem [{\citenamefont {Smith}\ \emph {et~al.}(2001)\citenamefont {Smith},
  \citenamefont {Tans}, \citenamefont {Smith}, \citenamefont {Grimes},
  \citenamefont {Anderson},\ and\ \citenamefont {Bustamante}}]{smith01}%
  \BibitemOpen
  \bibfield  {author} {\bibinfo {author} {\bibfnamefont {D.~E.}\ \bibnamefont
  {Smith}}, \bibinfo {author} {\bibfnamefont {S.~J.}\ \bibnamefont {Tans}},
  \bibinfo {author} {\bibfnamefont {S.~B.}\ \bibnamefont {Smith}}, \bibinfo
  {author} {\bibfnamefont {S.}~\bibnamefont {Grimes}}, \bibinfo {author}
  {\bibfnamefont {D.~L.}\ \bibnamefont {Anderson}}, \ and\ \bibinfo {author}
  {\bibfnamefont {C.}~\bibnamefont {Bustamante}},\ }\href@noop {} {\bibfield
  {journal} {\bibinfo  {journal} {Nature (London)},\ }\textbf {\bibinfo
  {volume} {413}},\ \bibinfo {pages} {748} (\bibinfo {year}
  {2001})}\BibitemShut {NoStop}%
\bibitem [{\citenamefont {Wen}\ \emph {et~al.}(2007)\citenamefont {Wen},
  \citenamefont {Manosas}, \citenamefont {Li}, \citenamefont {Smith},
  \citenamefont {Bustamante}, \citenamefont {Ritort},\ and\ \citenamefont
  {Tinoco}}]{wen07}%
  \BibitemOpen
  \bibfield  {author} {\bibinfo {author} {\bibfnamefont {J.-D.}\ \bibnamefont
  {Wen}}, \bibinfo {author} {\bibfnamefont {M.}~\bibnamefont {Manosas}},
  \bibinfo {author} {\bibfnamefont {P.~T.~X.}\ \bibnamefont {Li}}, \bibinfo
  {author} {\bibfnamefont {S.~B.}\ \bibnamefont {Smith}}, \bibinfo {author}
  {\bibfnamefont {C.}~\bibnamefont {Bustamante}}, \bibinfo {author}
  {\bibfnamefont {F.}~\bibnamefont {Ritort}}, \ and\ \bibinfo {author}
  {\bibfnamefont {I.}~\bibnamefont {Tinoco}},\ }\href@noop {} {\bibfield
  {journal} {\bibinfo  {journal} {Biophys. J.},\ }\textbf {\bibinfo {volume}
  {92}},\ \bibinfo {pages} {2996} (\bibinfo {year} {2007})}\BibitemShut
  {NoStop}%
\bibitem [{\citenamefont {Hosokawa}\ \emph {et~al.}(2005)\citenamefont
  {Hosokawa}, \citenamefont {Yoshikawa},\ and\ \citenamefont
  {Masuhara}}]{Hoso05}%
  \BibitemOpen
  \bibfield  {author} {\bibinfo {author} {\bibfnamefont {C.}~\bibnamefont
  {Hosokawa}}, \bibinfo {author} {\bibfnamefont {H.}~\bibnamefont {Yoshikawa}},
  \ and\ \bibinfo {author} {\bibfnamefont {H.}~\bibnamefont {Masuhara}},\
  }\href@noop {} {\bibfield  {journal} {\bibinfo  {journal} {Phys. Rev. E},\
  }\textbf {\bibinfo {volume} {72}},\ \bibinfo {pages} {021408} (\bibinfo
  {year} {2005})}\BibitemShut {NoStop}%
\bibitem [{\citenamefont {Oldham}(1983)}]{Old83}%
  \BibitemOpen
  \bibfield  {author} {\bibinfo {author} {\bibfnamefont {R.}~\bibnamefont
  {Oldham}},\ }\href@noop {} {\bibfield  {journal} {\bibinfo  {journal} {Cancer
  Metastasis Rev.},\ }\textbf {\bibinfo {volume} {2}},\ \bibinfo {pages} {323}
  (\bibinfo {year} {1983})}\BibitemShut {NoStop}%
\bibitem [{\citenamefont {Dholakia}\ and\ \citenamefont
  {Zemanek}(2010)}]{Dhol10}%
  \BibitemOpen
  \bibfield  {author} {\bibinfo {author} {\bibfnamefont {K.}~\bibnamefont
  {Dholakia}}\ and\ \bibinfo {author} {\bibfnamefont {P.}~\bibnamefont
  {Zemanek}},\ }\href@noop {} {\bibfield  {journal} {\bibinfo  {journal} {Rev.
  Mod. Phys.},\ }\textbf {\bibinfo {volume} {82}},\ \bibinfo {pages} {595}
  (\bibinfo {year} {2010})}\BibitemShut {NoStop}%
\bibitem [{\citenamefont {Padget}\ and\ \citenamefont {Bowman}(2011)}]{Pad11}%
  \BibitemOpen
  \bibfield  {author} {\bibinfo {author} {\bibfnamefont {M.}~\bibnamefont
  {Padget}}\ and\ \bibinfo {author} {\bibfnamefont {R.}~\bibnamefont
  {Bowman}},\ }\href@noop {} {\bibfield  {journal} {\bibinfo  {journal} {Nature
  Photon.},\ }\textbf {\bibinfo {volume} {5}},\ \bibinfo {pages} {343}
  (\bibinfo {year} {2011})}\BibitemShut {NoStop}%
\bibitem [{\citenamefont {Lee}\ and\ \citenamefont {Chang}(2006)}]{Lee06}%
  \BibitemOpen
  \bibfield  {author} {\bibinfo {author} {\bibfnamefont {H.~B. W.~G.}\
  \bibnamefont {Lee}}\ and\ \bibinfo {author} {\bibfnamefont {J.~K.}\
  \bibnamefont {Chang}},\ }\href@noop {} {\bibfield  {journal} {\bibinfo
  {journal} {Curr. Appl. Phys.},\ }\textbf {\bibinfo {volume} {6S1}},\ \bibinfo
  {pages} {e237} (\bibinfo {year} {(2006)})}\BibitemShut {NoStop}%
\bibitem [{\citenamefont {Ahlawat}\ \emph {et~al.}(2007)\citenamefont
  {Ahlawat}, \citenamefont {Dasgupta},\ and\ \citenamefont {Gupta}}]{Ahl07}%
  \BibitemOpen
  \bibfield  {author} {\bibinfo {author} {\bibfnamefont {S.}~\bibnamefont
  {Ahlawat}}, \bibinfo {author} {\bibfnamefont {R.}~\bibnamefont {Dasgupta}}, \
  and\ \bibinfo {author} {\bibfnamefont {P.~K.}\ \bibnamefont {Gupta}},\
  }\href@noop {} {\bibfield  {journal} {\bibinfo  {journal} {J. Opt. A: Pure
  Appl. Opt},\ }\textbf {\bibinfo {volume} {9}},\ \bibinfo {pages} {S189}
  (\bibinfo {year} {(2007)})}\BibitemShut {NoStop}%
\bibitem [{\citenamefont {Berg-Sorensen}\ and\ \citenamefont
  {Flyvbjerg}(2004)}]{berg04}%
  \BibitemOpen
  \bibfield  {author} {\bibinfo {author} {\bibfnamefont {K.}~\bibnamefont
  {Berg-Sorensen}}\ and\ \bibinfo {author} {\bibfnamefont {H.}~\bibnamefont
  {Flyvbjerg}},\ }\href@noop {} {\bibfield  {journal} {\bibinfo  {journal}
  {Rev. Sci. Inst.},\ }\textbf {\bibinfo {volume} {75}} (\bibinfo {year}
  {2004})}\BibitemShut {NoStop}%
\bibitem [{\citenamefont {Macdonald}\ and\ \citenamefont {Paterson}\ and\ \citenamefont {Sibett}\ and\ \citenamefont {Dholakia}\ and\ \citenamefont{Bryant}(2001)}]{Mac01}%
  \BibitemOpen
  \bibfield  {author} {\bibinfo {author} {\bibfnamefont {M.~P.}~\bibnamefont
  {Macdonald}}, \bibinfo {author} {\bibfnamefont {L.}~\bibnamefont
  {Paterson}}, \bibinfo {author} {\bibfnamefont {W.}~\bibnamefont
  {Sibett}}, \bibinfo {author} {\bibfnamefont {K.}~\bibnamefont
  {Dholakia}}, \ and\ \bibinfo {author} {\bibfnamefont {P.~E.}~\bibnamefont
  {Bryant}},\ }\href@noop {} {\bibfield  {journal} {\bibinfo  {journal} {Opt. Lett.},\ }\textbf {\bibinfo {volume} {26}},\ \bibinfo {pages}
  {863} (\bibinfo {year}
  {2001})}\BibitemShut {NoStop}%
\bibitem [{\citenamefont {Neuman}\ and\ \citenamefont {Block}(2004)}]{Neu04}%
  \BibitemOpen
  \bibfield  {author} {\bibinfo {author} {\bibfnamefont {K.~C.}\ \bibnamefont
  {Neuman}}\ and\ \bibinfo {author} {\bibfnamefont {S.~M.}\ \bibnamefont
  {Block}},\ }\href@noop {} {\bibfield  {journal} {\bibinfo  {journal} {Rev. of
  Sci. Inst.},\ }\textbf {\bibinfo {volume} {75}}, \bibinfo {pages}
  {2787} (\bibinfo {year}
  {2004})}\BibitemShut {NoStop}%
\bibitem [{\citenamefont {Sheppard}\ and\ \citenamefont {Gu}(1992)}]{Mgu92}%
  \BibitemOpen
  \bibfield  {author} {\bibinfo {author} {\bibfnamefont {C.~J.~R.}\
  \bibnamefont {Sheppard}}\ and\ \bibinfo {author} {\bibfnamefont
  {M.}~\bibnamefont {Gu}},\ }\href@noop {} {\bibfield  {journal} {\bibinfo
  {journal} {Opt. Comm.},\ }\textbf {\bibinfo {volume} {41}},\ \bibinfo {pages}
  {180} (\bibinfo {year} {1992})}\BibitemShut {NoStop}%
\bibitem [{\citenamefont {Neves}\ \emph {et~al.}(2007)\citenamefont {Neves},
  \citenamefont {Fontes}, \citenamefont {Cesar}, \citenamefont {Camposeo},
  \citenamefont {Cingolani},\ and\ \citenamefont {Pisignano}}]{AAR07}%
  \BibitemOpen
  \bibfield  {author} {\bibinfo {author} {\bibfnamefont {A.~A.~R.}\
  \bibnamefont {Neves}}, \bibinfo {author} {\bibfnamefont {A.}~\bibnamefont
  {Fontes}}, \bibinfo {author} {\bibfnamefont {C.~L.}\ \bibnamefont {Cesar}},
  \bibinfo {author} {\bibfnamefont {A.}~\bibnamefont {Camposeo}}, \bibinfo
  {author} {\bibfnamefont {R.}~\bibnamefont {Cingolani}}, \ and\ \bibinfo
  {author} {\bibfnamefont {D.}~\bibnamefont {Pisignano}},\ }\href@noop {}
  {\bibfield  {journal} {\bibinfo  {journal} {Phys. Rev. E},\ }\textbf
  {\bibinfo {volume} {76}},\ \bibinfo {pages} {061917} (\bibinfo {year}
  {(2007)})}\BibitemShut {NoStop}%
\bibitem [{\citenamefont {Born}\ and\ \citenamefont {Wolf}(1989)}]{Born}%
  \BibitemOpen
  \bibfield  {author} {\bibinfo {author} {\bibfnamefont {M.}~\bibnamefont
  {Born}}\ and\ \bibinfo {author} {\bibfnamefont {E.}~\bibnamefont {Wolf}},\
  }\href@noop {} {\emph {\bibinfo {title} {Principles of Optics}}}\ (\bibinfo
  {publisher} {Pergamon Press},\ \bibinfo {year} {1989})\BibitemShut {NoStop}%
\bibitem [{\citenamefont {Rohrbach}(2005)}]{Roh05}%
  \BibitemOpen
  \bibfield  {author} {\bibinfo {author} {\bibfnamefont {A.}~\bibnamefont
  {Rohrbach}},\ }\href@noop {} {\bibfield  {journal} {\bibinfo  {journal}
  {Phys. Rev. Lett.},\ }\textbf {\bibinfo {volume} {95}},\ \bibinfo {pages}
  {168102} (\bibinfo {year} {2005})}\BibitemShut {NoStop}%
\bibitem [{\citenamefont {Zem\'{a}nek}\ \emph {et~al.}(2001)\citenamefont
  {Zemánek}, \citenamefont {Jonás},\ and\ \citenamefont {Florin}}]{Zem01}%
  \BibitemOpen
  \bibfield  {author} {\bibinfo {author} {\bibfnamefont {P.}~\bibnamefont
  {Zemánek}}, \bibinfo {author} {\bibfnamefont {A.}~\bibnamefont {Jonás}}, \
  and\ \bibinfo {author} {\bibfnamefont {E.-L.}\ \bibnamefont {Florin}},\
  }\href@noop {} {\bibfield  {journal} {\bibinfo  {journal} {Opt. Lett.},\
  }\textbf {\bibinfo {volume} {26}},\ \bibinfo {pages} {1466} (\bibinfo {year}
  {2001})}\BibitemShut {NoStop}%
\bibitem [{\citenamefont {McCann}\ \emph {et~al.}(1999)\citenamefont {McCann},
  \citenamefont {Dykman},\ and\ \citenamefont {Golding}}]{Mcc99}%
  \BibitemOpen
  \bibfield  {author} {\bibinfo {author} {\bibfnamefont {L.~I.}\ \bibnamefont
  {McCann}}, \bibinfo {author} {\bibfnamefont {M.}~\bibnamefont {Dykman}}, \
  and\ \bibinfo {author} {\bibfnamefont {B.}~\bibnamefont {Golding}},\
  }\href@noop {} {\bibfield  {journal} {\bibinfo  {journal} {Nature},\ }\textbf
  {\bibinfo {volume} {402}},\ \bibinfo {pages} {785} (\bibinfo {year}
  {1999})}\BibitemShut {NoStop}%
\bibitem [{\citenamefont {Malagnino}\ \emph {et~al.}(2002)\citenamefont
  {Malagnino}, \citenamefont {Pescea}, \citenamefont {Sasso},\ and\
  \citenamefont {Arimondo}}]{Gou88}%
  \BibitemOpen
  \bibfield  {author} {\bibinfo {author} {\bibfnamefont {N.}~\bibnamefont
  {Malagnino}}, \bibinfo {author} {\bibfnamefont {G.}~\bibnamefont {Pescea}},
  \bibinfo {author} {\bibfnamefont {A.}~\bibnamefont {Sasso}}, \ and\ \bibinfo
  {author} {\bibfnamefont {E.}~\bibnamefont {Arimondo}},\ }\href@noop {}
  {\bibfield  {journal} {\bibinfo  {journal} {Opt. Comm.},\ }\textbf {\bibinfo
  {volume} {214}},\ \bibinfo {pages} {15} (\bibinfo {year} {2002})}\BibitemShut
  {NoStop}%
\bibitem [{\citenamefont {Zhang}\ \emph {et~al.}(2006)\citenamefont {Zhang},
  \citenamefont {Gu}, \citenamefont {Schwartzberg}, \citenamefont {Chen},\ and\
  \citenamefont {Zhang}}]{Zhan06}%
  \BibitemOpen
  \bibfield  {author} {\bibinfo {author} {\bibfnamefont {Y.}~\bibnamefont
  {Zhang}}, \bibinfo {author} {\bibfnamefont {C.}~\bibnamefont {Gu}}, \bibinfo
  {author} {\bibfnamefont {A.~M.}\ \bibnamefont {Schwartzberg}}, \bibinfo
  {author} {\bibfnamefont {S.}~\bibnamefont {Chen}}, \ and\ \bibinfo {author}
  {\bibfnamefont {J.~Z.}\ \bibnamefont {Zhang}},\ }\href@noop {} {\bibfield
  {journal} {\bibinfo  {journal} {Phys. Rev. B},\ }\textbf {\bibinfo {volume}
  {73}},\ \bibinfo {pages} {165405} (\bibinfo {year} {2006})}\BibitemShut
  {NoStop}%
\bibitem [{\citenamefont {Grier}(2003)}]{Gri03}%
  \BibitemOpen
  \bibfield  {author} {\bibinfo {author} {\bibfnamefont {D.}~\bibnamefont
  {Grier}},\ }\href@noop {} {\bibfield  {journal} {\bibinfo  {journal}
  {Nature},\ }\textbf {\bibinfo {volume} {424}},\ \bibinfo {pages} {810}
  (\bibinfo {year} {2003})}\BibitemShut {NoStop}%
\bibitem [{\citenamefont {Brzobohat/'{y}}\ \emph {et~al.}(2010)\citenamefont {Brzobohat/'{y}},
  \citenamefont {\u{C}i\u{z}im\'{a}r}, \citenamefont {Kar\'{a}sek}, \citenamefont {\u{S}iler},
  \citenamefont {Dholakia},\ and\ \citenamefont {Zem\'{a}nek}}]{Brz10}%
  \BibitemOpen
  \bibfield  {author} {\bibinfo {author} {\bibfnamefont {O.}\
  \bibnamefont {Brzobohat\'{y}}}, \bibinfo {author} {\bibfnamefont {T.}~\bibnamefont  {\u{C}i\u{z}im\'{a}r}}, \bibinfo {author} {\bibfnamefont {V.}\ \bibnamefont {Kar\'{a}sek}},  \bibinfo {author} {\bibfnamefont {M.}~\bibnamefont {\u{S}iler}}, \bibinfo
  {author} {\bibfnamefont {K.}~\bibnamefont {Dholakia}}, \ and\ \bibinfo
  {author} {\bibfnamefont {P.}~\bibnamefont {Zem\'{a}nek}},\ }\href@noop {}
  {\bibfield  {journal} {\bibinfo  {journal} {Opt. Exp.},\ }\textbf
  {\bibinfo {volume} {18}},\ \bibinfo {pages} {25389} (\bibinfo {year}
  {(2010)})}\BibitemShut {NoStop}%

\end{thebibliography}
%

\end{document}